\documentclass[journal,draftclsnofoot,onecolumn,12pt]{IEEEtran}

\usepackage{filecontents}
\usepackage{glossaries}
\usepackage[table,xcdraw]{xcolor}
\usepackage{amsthm}
\usepackage{epsfig}
\usepackage{times}
\usepackage{verbatim}
\usepackage{enumerate}
\usepackage{multicol}
\usepackage{afterpage}
\usepackage{wrapfig}
\usepackage{cite}
\usepackage{graphicx}
\usepackage{algorithm}
\usepackage{algpseudocode}
\usepackage{ifmtarg}
\usepackage{amssymb}
\usepackage{amsmath}
\usepackage{color}
\usepackage{bbm}
\usepackage{bm}
\usepackage{epstopdf}
\usepackage{flushend}
\usepackage{xspace}
\usepackage{caption}
\usepackage[caption=false,font=footnotesize]{subfig}
 \usepackage{multirow}
 \usepackage{stackengine}

\theoremstyle{remark}

\def\bb0{{\mathbb{0}}}

\def\bbE{{\mathbb{E}}}

\def\bbP{{\mathbb{P}}}

\def\ba{{\mathbf{a}}}
\def\bb{{\mathbf{b}}}
\def\bc{{\mathbf{c}}}
\def\bd{{\mathbf{d}}}

\def\bs{{\mathbf{s}}}

\def\bz{{\mathbf{z}}}
\def\b0{{\mathbf{0}}}

\def\bA{{\mathbf{A}}}

\def\bX{{\mathbf{X}}}

\def\cA{\mathcal{A}}
\def\cB{\mathcal{B}}
\def\cC{\mathcal{C}}

\def\cL{\mathcal{L}}

\def\cR{\mathcal{R}}
\def\cS{\mathcal{S}}
\def\cT{\mathcal{T}}
\def\cU{\mathcal{U}}

\def\cX{\mathcal{X}}

\def\sf0{{\mathsf{0}}}

\newcommand{\figref}[1]{Fig.~\ref{#1}}
\newcommand{\tabref}[1]{Table~\ref{#1}}
\newcommand{\algoref}[1]{Algorithm~\ref{#1}}

\newcommand{\secref}[1]{Section~\ref{#1}}

\newcommand{\doublebars}[1]{\left\Vert#1\right\Vert}

\newcommand{\msinr}{\mathrm{SINR}}

\newcommand{\tho}[1]{T_{\mathrm{HO},#1}}

\newcommand{\indicator}[1]{\mathbbm{1}_{#1}}
\newcommand{\st}{\mathrm{subject~to}}
\newcommand{\maxcap}{max-$C_{\mathrm{Shannon}}$\xspace}

\newcommand{\BS}{\gls{bs}\xspace}
\newcommand{\BSs}{\glspl{bs}\xspace}
\newcommand{\snr}{\gls{snr}\xspace}
\newcommand{\sinr}{\gls{sinr}\xspace}

\newcommand{\mpc}{\gls{mpc}\xspace}
\newcommand{\mdp}{\gls{mdp}\xspace}
\newcommand{\rl}{\gls{rl}\xspace}

\newcommand{\ho}{handover\xspace}
\newcommand{\hos}{handovers\xspace}

\newcommand{\hit}{\gls{hit}\xspace}
\newcommand{\cdf}{\gls{cdf}\xspace}

\newcommand{\lstm}{\gls{lstm}\xspace}
\newcommand{\nmse}{\gls{nmse}\xspace}
\newcommand{\ndcg}{\gls{ndcg}\xspace}
\newcommand{\dcg}{\gls{dcg}\xspace}
\newcommand{\idcg}{\gls{idcg}\xspace}

\newacronym{bs}{BS}{base station}
\newacronym{snr}{SNR}{signal-to-noise ratio}
\newacronym{sinr}{SINR}{signal-to-interference-plus-noise ratio}
\newacronym{inr}{INR}{interference-to-noise ratio}
\newacronym{rinr}{RINR}{residual-interference-to-noise ratio}
\newacronym{los}{LOS}{line-of-sight}
\newacronym{nlos}{NLOS}{non-line-of-sight}
\newacronym{cdf}{CDF}{cumulative density function}
\newacronym{pdf}{PDF}{probability density function}
\newacronym{fd}{FD}{full-duplex}
\newacronym{hd}{HD}{half-duplex}
\newacronym{si}{SI}{self-interference}
\newacronym{ho}{HO}{handover}
\newacronym{hit}{HIT}{handover interruption time}
\newacronym{mmw}{mmWave}{millimeter wave}
\newacronym{fr2}{FR2}{frequency range 2}
\newacronym{iab}{IAB}{integrated access and backhaul}
\newacronym{mpc}{MPC}{model predictive control}
\newacronym{mdp}{MDP}{Markov decision process}
\newacronym{rl}{RL}{reinforcement learning}
\newacronym{ula}{ULA}{uniform linear array}
\newacronym{rf}{RF}{radio frequency}
\newacronym{dft}{DFT}{discrete fourier transform}
\newacronym{lstm}{LSTM}{long short-term memory}
\newacronym{nmse}{NMSE}{normalized mean-squared error}
\newacronym{ndcg}{NDCG}{normalized discounted cumulative gain}
\newacronym{idcg}{IDCG}{ideal DCG}
\newacronym{dcg}{DCG}{discounted cumulative gain}

\newcommand{\mycomment}[1]{}
\usepackage{mathtools} 

\makeatother
\begin{document}
\title{Forecaster-aided User Association and Load Balancing in Multi-band Mobile Networks}

	\author{
		\IEEEauthorblockN{\large  Manan Gupta, Sandeep Chinchali, Paul Varkey, and Jeffrey~G.~Andrews}\\
		\thanks{M. Gupta, S. Chinchali, and J. G. Andrews are with 6G@UT in the Wireless Networking and Communications Group at the University of Texas at Austin. P. Varkey is a Staff Engineer with Meta. 
        }
  }
	
	\maketitle
	\begin{abstract}
Cellular networks are becoming increasingly heterogeneous with higher \BS densities and ever more frequency bands, making \BS selection and band assignment key decisions in terms of rate and coverage.
In this paper, we decompose the mobility-aware user association task into (i) forecasting of user rate and then (ii) convex utility maximization for user association accounting for the effects of \BS load and \ho overheads. 
Using a linear combination of \nmse and \ndcg as a novel loss function, a recurrent deep neural network is trained to reliably forecast the mobile users' future rates.
Based on the forecast, the controller optimizes the association decisions to maximize the service rate-based network utility using our computationally efficient (speed up of $100\times$ versus generic convex solver) algorithm based on the Frank-Wolfe method.
Using an industry-grade network simulator developed by Meta, we show that the proposed \mpc approach improves the $5$th percentile service rate by $3.5\times$ compared to the traditional signal strength-based association, reduces the median number of \hos by $7\times$ compared to a \ho agnostic strategy, and achieves service rates close to a genie-aided scheme.
Furthermore, our model-based approach is significantly more sample-efficient (needs $100\times$ less training data) compared to model-free \rl, and generalizes well across different user drop scenarios.
	\end{abstract}

\glsresetall
\section{Introduction}
\label{Sec:Intro}
Cellular networks continue to trend towards multi-tier and multi-band deployments to meet intense consumer demands for faster and more ubiquitous data connectivity.
The use of low-power small cell technology enables flexible deployments and helps boost network capacity through increased spatial reuse.
Meanwhile, more spectrum is being released, for instance almost 500 MHz recently in the 3.7 - 4.2 GHz ``C-band" and more to be auctioned in the ``beachfront'' bands from 1 to 2.6 GHz.  Assigning users to bands and cells is an increasingly complicated problem, with large implications on the overall network coverage and throughput, particularly in the context of mobility.  The conventional strategy of simply maximizing \snr or \sinr for user association tends to lead to an imbalanced load distribution and can severely degrade both a given user's as well as the overall service rate \cite{Seven_Andrews13}.  Our goal in this paper is to develop an optimal band assignment and \BS selection strategy for multi-band heterogeneous networks that accounts for the bandwidth and propagation characteristics of each band, along with the traffic load,  user mobility, and handover overhead.

\subsection{Motivation, Background, and Scope of Paper}
A multi-band deployment increases the overall bandwidth as well as number of connection options for subscribers, however, it comes at the cost of increased complexity, especially in the context of mobility \cite{andrews14overview}.
Traditional \emph{user association}---meaning a user's \BS selection and band assignment---policies that maximize received signal power or \sinr lead to preferential association towards the lowest frequency bands---which offer better penetration and lower pathloss---and to tower-mounted macrocells with higher transmit power.
Such association rules are highly sub-optimal in terms of throughput, because higher carrier frequencies typically offer wider bandwidths while small cells are much more lightly loaded.
Furthermore, heterogeneous deployments with small coverage areas are more prone to frequent \hos and consequently poor service rates due to transmission delays, ping-pong effects, and connection drops \cite{Arshad_Velocity17, feriani22multiobjective}. 

As a result, there has been considerable work investigating user association and cell load balancing.
These include stochastic geometry studies which analytically characterize the throughput gain from load balancing \cite{singh13offloading, kassir22analysis, Modeling_Lin13} and optimization-based user association strategies \cite{shen14distributed, ye13user, kim12distributed, ye16user, liu19joint, teng21joint}.
In \cite{ye13user, shen14distributed}, authors propose log-utility-based association strategies to achieve network-wide proportional fairness for static users, and \cite{kim12distributed} presents $\alpha$-optimal load balancing strategy for inhomogeneous user traffic.
The user association problem has also been studied in conjunction with power allocation \cite{liu19joint}, massive MIMO \cite{ye16user}, and \BS activation strategies \cite{teng21joint}, under the network utility maximization framework.

Accounting for user mobility within the optimization framework is generally intractable, with limited studies in the literature, mostly using simplistic mobility models.
However, a variety of studies underscore the benefits of mobility-aware user association and \ho management \cite{addali19dynamic, lin13towards, hsueh17equivalent, Arshad_Velocity17}.
The authors in \cite{lin13towards, hsueh17equivalent} characterize \ho rate as a function of user velocity and \BS density. 
By leveraging similar stochastic geometry-based analysis, \cite{Arshad_Velocity17} proposes a \BS skipping \ho policy for high-speed users in a two-tier cellular network.
For a comprehensive survey of mobility models---both trace-based and probabilistic---and mobility-aware performance characterizations, please refer to \cite{tabassum19fundamentals}.

Meanwhile, in actual cellular deployments, proactive load balancing is usually accomplished by biasing user association towards small cells \cite{rakotomanana16optimum, ghosh12heterogeneous} by some tunable dB value.
For example, a small cell could be assigned a 5-10 dB \snr bias relative to a macrocell \cite{QualcommHeNets}. 
However, such rule-based strategies are not only suboptimal, but they require carefully tuning the bias values and struggle to cope with time-varying traffic patterns from mobile users and for large number of connections \cite{lin22embracing}.
As the industry trends towards intelligent and open networks \cite{polese22understanding, lin22embracing}, network measurements across \BSs and users can be leveraged to make data-driven user association and load balancing decisions.
As such, mobility-aware user association has been considered recently as a \rl problem \cite{zhang18loadbalancing, Handover_Wang18, chinchali18cellular, zhao19deep, xu19load, gupta21load, wu21load, feriani22multiobjective, lacava22programmable}.
Among these, \cite{zhao19deep, gupta21load} aim to optimize the network utility within the \rl framework, \cite{zhang18loadbalancing, Handover_Wang18} maximize the sum-rate, and \cite{xu19load, feriani22multiobjective} optimize cell individual offset parameters.
However, such model-free \rl techniques are known to be sample-inefficient \cite{mai2022sample}, and there has been a push towards scalable and sample-efficient formulations \cite{wu21load, alcaraz22model,agarwal2022taskdriven}. 
The authors in \cite{wu21load} use transfer learning and a model-based \rl solution was explored in \cite{alcaraz22model} for load balancing and network optimization.

In this work, we focus on the most widely deployed frequency bands, namely the ``sub-6 GHz" or ``frequency range 1 (FR1)" range, which still carry the vast majority of cellular traffic even in 5G deployments today, despite the introduction of millimeter wave bands. 
At these lower frequencies, initial access and control channel communication can be reliably performed without the requirement for highly directional beamforming, which greatly simplifies the user association and \ho decisions.
A key challenge for load balancing with highly directional transmissions is that it leads to directional and time-varying interference, load-based association decisions problematic and non-convex.
Furthermore, since such bands will be lightly loaded for the forseeable future, it may not be necessary to perform load balancing on them.
Thus, we restrict our attention to the lower bands in this work.

\subsection{Contributions}

The overarching technical contribution of this paper is the development of a novel framework for mobility-aware \BS and frequency band association in a large-scale cellular network.  The key elements and unique aspects of our approach are summarized as follows.

\textbf{Optimization framework for mobility and \ho aware association.}
We formulate a novel optimization problem to maximize network utility based on the users' service rates.
In \secref{subsec:service_rate} we define the user service rate, which depends on the load on the serving \BS, the \sinr between the user and \BS, and the signaling overhead in case of a triggered \ho.
A key insight is that in a mobile network, the user association task can be decomposed into forecasting user trajectories and thus their future rates, and then a convex utility maximization problem based on those predicted rate values.
As such, we propose a learning-based \mpc approach in \secref{sec:mpc_formulation}, where we only need data-driven learning for forecasting, while we can employ interpretable utility maximization for band assignment and \BS selection.

\textbf{Sample-efficient forecaster design.} 
In \secref{sec:forecaster_design}, we design a forecaster to model and predict the users' mobility patterns, which are then used by the \mpc controller to solve for associations that maximize the network utility.
The forecaster is trained in a supervised fashion, which is generally more sample-efficient and stable compared to model-free \rl.
To further improve the sample-efficiency, in we identify important characteristics of a ``good" forecast for the given utility maximization problem and develop a novel loss function to train a task-aware forecaster, instead of using the task-agnostic mean-squared error loss.
We compare the forecaster-based approach with \rl in terms of generalization ability and sample-efficiency and show that the proposed approach can learn the system dynamics induced by user mobility with $100\times$ less training samples.

\textbf{Computationally efficient optimization algorithm.}
In \secref{sec:frank_wolfe_algo}, we propose an optimization algorithm to quickly solve the constrained utility maximization problem for user association.
We exploit the structure of the constraint set to derive an analytical solution for the direction-finding step in the projection-free Frank-Wolfe method.
Leveraging the analytical solution and backtracking line-search to dynamically adjust the step-size in every iteration, the execution time for solving the utility maximization can be reduced by two orders of magnitude compared to a generic convex solver.

\textbf{Realistic validation of rate gain and \ho reduction.}
In \secref{sec:results}, we evaluate the performance of the proposed forecaster-aided \mpc solution through extensive simulations using an industry-grade network simulator developed by Meta (formerly Facebook) that accurately models a commercial cellular network.  We demonstrate that our \mpc approach can provide a more uniform user experience and improve the 5th percentile service rate by about $3.5\times$ compared to conventional strategies.
Furthermore, we show that by using our forecaster-based mobility-aware user association strategy, the median number of \hos in a time slot can be simultaneously reduced by about $7\times$ compared to a \ho agnostic strategy.

\textit{Notation}: $\bA_{:,i}$ denotes the $i$-th column of matrix $\bA$, $\bA_{i,:}$ denotes the $i$-th row of matrix $\bA$, $\bA^T$ denotes the transpose of a matrix, $[\bA]_{i,j}$ denotes the $(i,j)$-th element of $\bA$, and $\bA^\dagger$ denotes the Hermitian transpose of a matrix. The $i$-th element of vector $\ba$ is denoted by $a_i$. $\indicator{\cA}$ denotes the indicator function over set $\cA$, and $|\cA|$ denotes the cardinality of set $\cA$.
\section{System Model}
\label{Sec:system_model}
\begin{figure}[t!]
    \centering
    \includegraphics[width=4in,                 keepaspectratio]{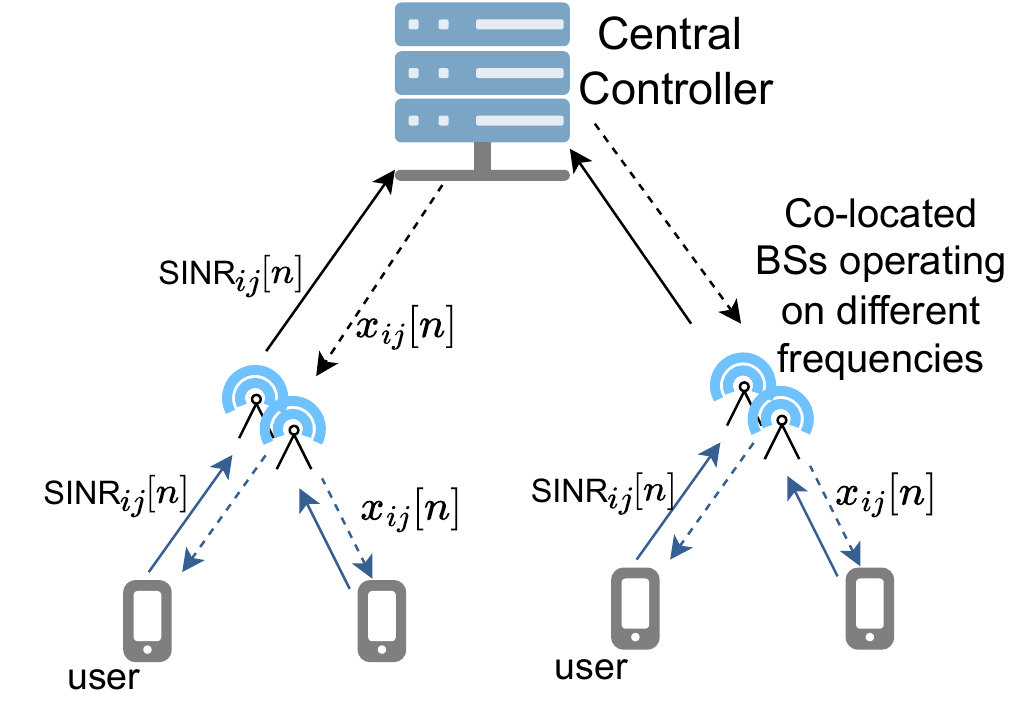}
    \caption{Network structure and flow of \sinr measurements. The \BSs receive \sinr measurements from the user in their cell. The central controller receives \sinr measurements from the \BSs and communicates the association indicators back to the \BSs and users.}
    \label{fig:network_struct}
\end{figure}

We consider a downlink cellular network which consists of $K$ different carrier frequency bands.
Let $\cB_k, \text{ where } k\in\{1, 2, \dots, K\}$, denote the set of all \BSs operating on the $k$-th frequency band and $\cU$ denote the set of all users in the network.
We characterize a \BS by its physical location \textit{and} its carrier frequency band. 
Thus, in this formulation, radio units operating on different carrier frequencies but mounted on the same physical tower are treated as different co-located \BSs, as shown in \figref{fig:network_struct}.
Note that $\cB_k$'s are disjoint sets and the set of all \BSs is denoted as $\cB = \bigcup_{k=1}^K\cB_k$.
As described later in \secref{sec:forecaster_design}, the set $\cU$ can vary over time.

We will assume slotted time indexed by $n$.
The users are mobile and each is initially associated with a parent \BS. 
It is assumed that user (re-)association decisions are made on a larger timescale compared to the scheduling of time and frequency resources.
Consequently, many scheduling slots are subsumed under a user association slot \cite{ye13user, shen14distributed}.
With this distinction, henceforth, a time slot will refer to a user association slot unless stated otherwise.
The \sinr used for association is averaged over the association interval, and thus, the small-scale fading is also averaged out.
The \sinr of the signal delivered to user $i$ by \BS $j\in\cB_k$ during time slot $n$ can be expressed as 
\begin{align}
    \msinr_{ij}[n] &= \frac{P_j g_{ij}[n]}{\sum_{j'\in\cB_k, j'\neq j}P_{j'}g_{ij'}[n] + W_j\sigma^2},
\end{align}
where $P_j$ denotes the transmit power at \BS $j$, $g_{ij}[n]$ represents the channel gain between user $i$ and \BS $j$ and includes the pathloss, shadowing, and the antenna gains, and $\sigma^2$ denotes the power spectral density of the thermal noise.
Since the \BS index $j$ also identifies its frequency band and different frequency bands can operate on different bandwidths, we denote by $W_j$ the bandwidth at \BS $j$.

At the beginning of a time slot $n$, each user measures the \sinr from its neighboring \BSs and reports these measurements to the parent \BS.
The \BSs then forward these \sinr measurements to a central network controller, which processes these measurements and computes the association indicators $x_{ij}[n]$, which are defined as 
\begin{align}
    x_{ij}[n] &= \begin{dcases}
                    1, & \textrm{user $i$ associates to \BS $j$ in slot $n$} \\
                    0, & \textrm{otherwise}.
                 \end{dcases}
\end{align}
These new association decisions are communicated to the \BSs by the controller and then forwarded to the users by their current parent \BS.
The remaining time in slot $n$ is then used for \ho and data transmission.
A \ho is triggered for user $i$ if $x_{ij}[n] \neq x_{ij}[n-1]$. 
Note that in our formulation, switching frequency bands on the same site also counts as a \ho.
\figref{fig:network_struct} summarizes the signal flow in a time slot.
Such closed-loop control is a major thrust for open radio access networks (O-RAN) where the RAN intelligent controller (RIC) can centrally control hundreds of \BSs while efficiently handling feedback and real-time centralized optimization \cite{lin22embracing, lacava22programmable}.
Furthermore, the central availability of data and analytics in the network enables using intelligent data-driven approaches to user association and generally network resource management by taking a user-centric approach \cite{polese22understanding}.

\section{Formulating Load Balancing as Utility Maximization}
\label{Sec:problem_formulation}

The system model discussed in the previous section accounts for multiple sources of heterogeneity. 
Each \BS could have a different transmit power $P_j$, the bandwidth $W_j$ can vary across \BSs thus affecting the receiver noise and the achievable rate as the pre-logarithmic term, and each \BS can operate on a different carrier frequency with vastly different propagation characteristics.
We now formulate a network utility maximization problem with the intention of fairly utilizing the available heterogeneous resources.
The metric of interest in our formulation is the user's \textit{service rate} which is a function of \sinr, \textit{load} on the serving \BS, and average throughput degradation experienced by the user due to a triggered \ho.
To ground intuition, we first present a genie-aided optimization which requires non-causal measurements.
In the next section, we propose a learning-based user association solution which leverages a forecaster, modeled as a deep neural network, to estimate these non-causal measurements.
In the following, we describe in detail the service rate and subsequently piece together the genie-aided utility maximization.

\subsection{User Service Rate}
\label{subsec:service_rate}
In a cellular network, each \BS typically serves multiple users and it must share its available time and frequency resources among these users. 
As a result, user's service rate depends on the load on its serving \BS and on the deployed resource scheduling algorithm.
Furthermore, while seamless \hos are necessary to support user mobility---a key feature of cellular networks---they involve relatively complicated signaling procedures and come with a costly overhead \cite{andrews14overview}.
Frequent \hos can lead to increased connection drop rates, higher transmission delays, and induce ping-pong effects \cite{feriani22multiobjective}.
These can be especially detrimental for heterogeneous deployments which typically have a smaller average cell size.
Hence, we define a user's service rate as a function of all the aforementioned factors.

The achievable data rate between user $i$ and \BS $j$ at time $n$ can be expressed as
\begin{align}
    \label{eq:achievable_rate_def}
    c_{ij}[n] &= W_j\log_2(1 + \msinr_{ij}[n]).
\end{align}
Let $\tho{i}[n]$ denote the \hit associated with the network's \ho procedure and $T_s$ denote the total slot duration.
Although not explicit from the notation, the \hit is a function of the source and target \BSs and the quality of channel between the user and the two \BSs.
Furthermore, \hit also depends on the proprietary \ho procedure implemented by the operator.
We define the service rate for user $i$ from \BS $j$ during slot $n$ as 
\begin{align}
    \label{eq:service_rate_def}
    r_{ij}[n] &= \frac{c_{ij}[n]}{\ell_j[n]}\left(1 - \frac{\tho{i}[n]}{T_s}\right),
\end{align}
where $\ell_j[n] = \sum_{i\in\cU}x_{ij}[n]$ represents the total number of users being served by \BS $j$---the load on \BS $j$---and $1/\ell_{j}[n]$ represents the fraction of resources allocated to each user by the scheduler at \BS $j$.
A round robin resource scheduler would achieve this exactly and a proportional fair scheduler would achieve this on average \cite{vishwanath02opportunistic}.
The multiplicative term $\left(1 - \frac{\tho{i}[n]}{T_s}\right)$ represents the fraction of time remaining for data transmission after \ho related signaling. 
$\tho{i}[n]$ is zero if user $i$ does not change association during slot $n$.
Note that one can easily include an \sinr gap from the Shannon capacity to account for discrete modulation with finite constellation size.


\subsection{Genie-Aided User Association}
\label{subsec:genie_aided_mpc}
We now formulate the genie-aided network utility maximization problem.
There is a considerable body of literature exploring different applications of various network utility functions such as network-wide resource allocation \cite{yigal07fairness, gupta22system}, network-wide max-min fairness \cite{rasek20joint}, logarithmic utility for balanced load distribution \cite{ye13user}, and $\alpha$-optimal user association \cite{kim12distributed}.
We use the logarithmic utility function which maximizes the product of service rates to achieve a healthy balance between network sum-rate and fairness and naturally achieves load balancing \cite{srikant2014comnets}.
The optimization takes as input the achievable rates $c_{ij}[n]$ and the \glspl{hit} $\tho{i}$, and yields association indicators $x_{ij}[n]$ which maximize the sum of the logarithm of the service rates $r_{ij}[n]$.

A genie-aided network utility maximization can thus be expressed as
\begin{subequations} \label{opt:util_max_genie}
\begin{align}
\underset{\{x_{ij}[n]\}}{\mathrm{maximize}} ~&\sum_{n=1}^{H}\sum_{j\in\cB}\sum_{i\in\cU} x_{ij}[n] \log\left(\frac{c_{ij}[n]}{\ell_j[n]}\cdot\left(1 - \frac{\tho{i}[n]}{T_s}\right)\right) \label{opt:util_max_genie_obj}\\
\st~\nonumber\\ 
& \sum_{j\in\cB}x_{ij}[n] = 1 \quad \forall i\in \cU \label{opt:util_max_genie_ue_cons}\\
& x_{ij}[n]\in\{0,1\} \quad \forall i\in \cU,\forall j\in\cB \label{opt:util_max_genie_indicator_cons}\\
& \ell_j[n] = \sum_{u\in\cU}x_{uj}[n] \quad \forall j\in\cB. \label{opt:util_max_genie_load_def} 
\end{align}
\end{subequations}
The constraint \eqref{opt:util_max_genie_ue_cons} and the binary integer constraint \eqref{opt:util_max_genie_indicator_cons} together ensure that each user $i$ is associated to exactly one \BS in time slot $n$.
Constraint \eqref{opt:util_max_genie_load_def} defines the total load on \BS $j$ in slot $n$ as the total number of users being served by $j$.
The achievable rates $c_{ij}[n]$ are related to the \sinr by Shannon's formula as shown in \eqref{eq:achievable_rate_def}.
Note that the \ho penalty $\left(1 - \frac{\tho{i}[n]}{T_s}\right)$, is a function of $x_{ij}[1], \dots, x_{ij}[n]$.
In order to account for this dependency, our objective is to maximize the logarithmic utility for a time horizon $H$.
Here, $n=1$ represents the current time slot and $n>1$ represents the future.
As a result, the above optimization is genie-aided because it is solved over a future time horizon assuming non-causal knowledge of $c_{ij}[n]$ and $\tho{i}[n]$.
Nonetheless, it serves as an upper bound on the achievable logarithmic utility.

Given the unknown system dynamics induced by user mobility and \ho procedures, and the \textit{cumulative utility maximization} objective, \eqref{opt:util_max_genie} is a good candidate for an \rl-based solution.
Although, a powerful technique, model-free \rl is known to be sample-inefficient \cite{wu21load, alcaraz22model}, and does not scale well as the  number of users and \BSs increase.
As a result, it often converges to highly sub-optimal policies for large-scale networks with \BSs and users densities close to realistic deployments.
We instead propose a learning-based \mpc approach where a forecaster is trained to model the system dynamics in a supervised fashion, which is relatively more sample-efficient and stable, and has been quite successful in other fields \cite{agarwal2022taskdriven}.
In \secref{sec:results}, we will compare the proposed \mpc approach to \rl in terms of their sample-efficiency for training and their generalization ability to different user drops.
\section{Learning-based Model Predictive Control}
\label{sec:mpc_formulation}
\begin{figure}[t!]
    \centering
    \includegraphics[width=6in,                 keepaspectratio]{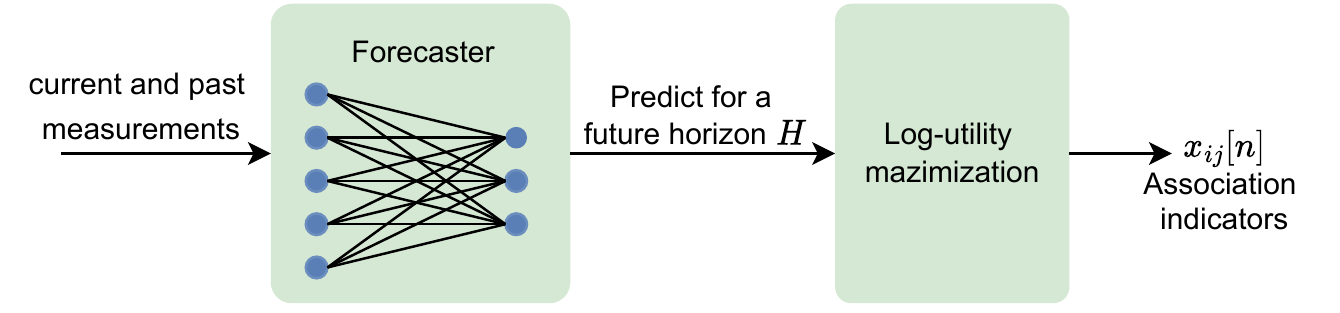}
    \caption{Forecaster-based MPC design. Current and past capacity measurements (computed from the \sinr's reported during earlier slots) are input to the forecaster which outputs a prediction of future capacities for a future horizon $H$. These are then used to generate new user associations which maximize the cumulative log-utility.}
    \label{fig:forecaster_mpc}
\end{figure}
The genie-aided optimization in \eqref{opt:util_max_genie} is non-causal because it requires future rate and \hit measurements, both of which are unknown at time $n$.
In this section, we propose a learning-based \mpc solution where we forecast system behavior and optimize the forecast to produce the best control decision at the current time \cite{mpc_james}.
For the considered mobile network,
the two main sources of uncertainty are (i) user mobility which affects both $c_{ij}[n]$ and \hit and (ii) the implemented \ho procedure which primarily affects \glspl{hit}.

The temporal variations in \sinr measurements---consequently in $c_{ij}[n]$---are largely a function of users' mobility patterns.
These are predictable to some extent, and for a reasonable time interval $T_s$ the next step in the trajectory is significantly correlated with the past trajectory \cite{liang03predictive}. 
Thus, to exploit these temporal correlations, we use a recurrent deep neural network architecture---specifically a \lstm---to forecast the future measurements using past and current samples.
We denote by $\hat{c}_{ij}[n]$ the prediction of $c_{ij}[n]$.

On the other hand, in addition to current channel conditions, the \hit also depends on the specific \ho procedure implemented by the operator.
Moreover, a \ho is not merely a physical layer task and involves control signal exchange across layers \cite{tayyab19survey}.
As a result, the \hit cannot be reliably predicted from only \sinr measurements and would require some degree of feedback from higher layers.
Thus, we instead model \hit as a random variable.

Distributing the logarithm over the product and using $\sum_{i}x_{ij}[n] = \ell_j[n]$, the objective \eqref{opt:util_max_genie_obj} can be written as 
\begin{align}
    \sum_{n=1}^{H}\sum_{j\in\cB}\sum_{i\in\cU} x_{ij}[n] \log(c_{ij}[n]) + x_{ij}[n]\log\left(1 - \frac{\tho{i}[n]}{T_s}\right) - \sum_{n=1}^H\sum_{j\in\cB}\ell_j[n]\log(\ell_j[n]).
\end{align}
The first term here is the logarithmic utility of the achievable rates, and as discussed earlier, we use a forecaster to predict these for $n>1$.
The third term is the entropy of the load variable $\ell_j[n]$ and is maximized when all \BSs are equally loaded.
The second term corresponds to the penalty for each \ho. Recall that $\tho{i}[n]$ is a function of the previous association of user $i$ as well as the new association, and is hard to reliably predict from physical layer \sinr measurements.
Thus, we model $\tho{i}[n]$ as a random variable and use the convex surrogate penalty
\begin{align}
    \label{eq:ho_surrogate_penalty}
    \frac{(x_{ij}[n] - x_{ij}[n-1])^2}{2}\bbE\left[\log\left(1 - \frac{\tho{i}[n]}{T_s}\right)\right].
\end{align}
Since, $x_{ij}[n]$ and $x_{ij}[n-1]$ are indicator variables, the quadratic term is $1$ if user $i$ goes through a \ho.
With these modifications, the user association \mpc can be expressed as 
\begin{subequations} \label{opt:util_max_mpc}
\begin{align}
\underset{\{x_{ij}[n]\}}{\mathrm{maximize}} ~&\sum_{n=1}^{H}\sum_{j\in\cB}\sum_{i\in\cU} x_{ij}[n] \log(\bar{c}_{ij}[n])  + \frac{\eta}{2}\left(x_{ij}[n] - x_{ij}[n-1]\right)^2 \nonumber\\
&-\sum_{n=1}^H\sum_{j\in\cB}\ell_j[n]\log(\ell_j[n])\label{opt:util_max_mpc_obj}\\
\st~\nonumber~\\
& \bar{c}_{ij}[n] = c_{ij}[n]\mathbbm{1}(n=1) + \hat{c}_{ij}[n]\mathbbm{1}(n>1) \label{opt:util_max_mpc_rate_def}\\
& \ell_j[n] = \sum_{u\in\cU}x_{uj}[n] \quad \forall j\in\cB \label{opt:util_max_mpc_load_def}\\ 
& \sum_{j\in\cB}x_{ij}[n] = 1 \quad \forall i\in \cU \label{opt:util_max_mpc_ue_cons}\\
& x_{ij}[n]\in\{0,1\} \quad \forall i\in \cU,\forall j\in\cB. \label{opt:util_max_mpc_indicator_cons}
\end{align}
\end{subequations}
Here, $\eta$ is the expected value defined in \eqref{eq:ho_surrogate_penalty} and is a negative constant. 
The association indicators $x_{ij}[0]$ refer to the user association from the previous time slot and are not optimization variables.
The above \mpc uses the current measurements for $n=1$ and the forecasts $\hat{c}_{ij}[n]$ for $n = 2, \dots, H$. 
The forecasting horizon is a hyperparameter which depends on the quality of the forecaster and the available compute power.
Typically, the reliability of the forecaster degrades as the prediction horizon increases, and as such larger $H$ can lead to undesirable control decisions.
The computational complexity of the \mpc increases dramatically with $H$ due to the quadratic \ho penalty; a major decision variable when choosing the right $H$.
The above \mpc has a concave objective---sum of an affine function, a negative quadratic, and entropy---with linear constraints and can be solved using a generic convex solver such as \cite{cvxpy1}.
We later present a variation of the Frank-Wolfe algorithm which can much more efficiently solve \eqref{opt:util_max_mpc}.

\section{Forecaster Design}
\label{sec:forecaster_design}
\begin{figure}[t!]
    \centering
    \includegraphics[width=6in,                 keepaspectratio]{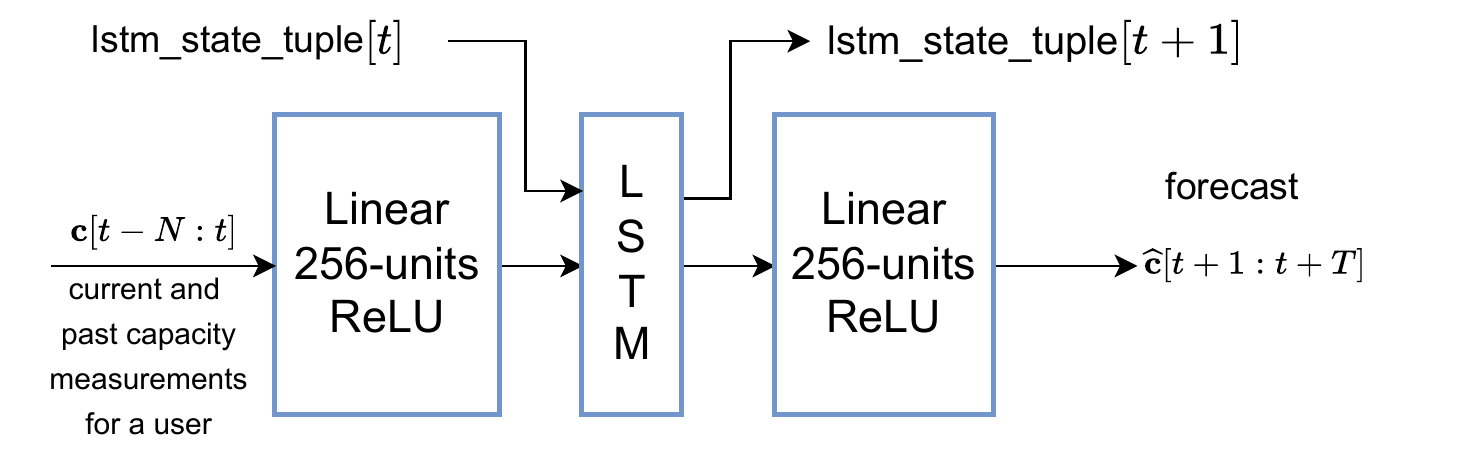}
    \caption{Neural network architecture for the forecaster.}
    \label{fig:forecaster_arch}
\end{figure}
Next, we describe the design of the forecaster which uses the time-series of current and past rates to predict the future measurements. 
Our goal is to use this forecast to solve for association indicators $x_{ij}[n]$ that maximize the network utility according to \eqref{opt:util_max_mpc}.
Hence, we design the forecaster to directly predict the achievable rates $c_{ij}[n]$, which are related to the \sinr measurements through \eqref{eq:achievable_rate_def}.
However, the task-aware loss function and the supervised learning procedure described in this section can be used to forecast any metric of interest.
For example, one could train an \sinr forecaster and then use \eqref{eq:achievable_rate_def} to get a prediction of the user rates if this better suits their particular use case.

We denote by $\bc_i[n]$ the $|\cB|\times 1$ vector of achievable rates of user $i$ from all \BSs in $\cB$.
To exploit the spatial and temporal correlations in $\bc_i[n]$, we use an \lstm-based architecture shown in \figref{fig:forecaster_arch}.
The input to the forecaster is the history of $N$ rate measurement vectors for a user $i$.
The forecaster then generates a rate prediction for $H$ future slots---also known as the forecasting horizon---denoted by $\hat{\bc}_i[n+1], \dots, \hat{\bc}_i[n+H]$.
This architecture has the following benefits.
First, the forecaster predicts the trajectory for each user individually and does not depend on the number of users in the network.
In fact, with this architecture, the learning-based \mpc formulation in \eqref{opt:util_max_mpc} can even support a time-varying $\cU$.
Second, during both the training and inference phases, measurements across users can be batched together and processed in parallel by the forecaster.

Since $\bc_i[n]$ are continuous variables, training the forecaster lies within the supervised regression paradigm, where the typical approach is to minimize the mean-squared error (MSE) between the prediction and true value.
However, our goal is to use the forecast from the trained model to solve for the load balancing user association indicators $x_{ij}[n]$.
Hence, we instead design a task-aware training loss function with the following two objectives in mind.
First, for a user $i$, the forecaster must preserve the \textit{ranking} induced by the capacity measurements.
Second, the forecaster must accurately predict the achievable rate between user $i$ and \BS $j$, relative to other \BSs.
In order words, it is satisfactory for the forecaster to be off, by say $1$ Mbps, as long as the forecast for every relevant \BS is off by approximately the same amount.
As such, we use the following loss function, which is a sum of the \nmse and \ndcg, to train the forecaster.
\begin{align}
    \label{eq:joint_loss}
    \cL(\bc_i, \hat{\bc}_i) &= \frac{\doublebars{\bc_i - \hat{\bc}_i}^2}{\doublebars{\bc_i}^2} + 1 - \mathrm{NDCG}(\bc_i, \hat{\bc}_i).
\end{align}
The first term in \eqref{eq:joint_loss} is the \nmse which penalizes any shift from the target values in the aforementioned relative sense, rather than in the absolute sense. 
The second term is the \ndcg ranking loss \cite{wang13theoretical} which is defined as
\begin{align}
    \label{eq:ndcg_def}
    \mathrm{NDCG}(\bc_i, \hat{\bc}_i) & \triangleq \frac{\mathrm{DCG}}{\mathrm{IDCG}}~~\textrm{where,}\\
    \mathrm{DCG} &= \sum_m \frac{g(\bc_{\hat{m}}) - 1}{\log_2(m)}\\
    \mathrm{IDCG} &= \sum_m \frac{g(\bc_{[m]}) - 1}{\log_2(m)}.
\end{align}
Here, $g(x) = 2^{\varsigma(x)}$, $\varsigma(\cdot)$ represents the sigmoid function, $(\cdot)_{[m]}$ denotes the $m$-th largest element of a vector and $\hat{m}$ is the index of the $m$-th largest element of the prediction $\hat{\bc}_i$.
The \ndcg metric is used in the ranking literature to quantify the performance of a recommendation system.
In such systems, it is more desirable to recommend most relevant results first where the relevance of each recommendation is measured by a \textit{relevance score}, which is given by $\bc_i$ for the load balancing task.
Unlike other ranking measures, \ndcg allows for a graded relevance score, and thus, is more suited for our use case.
\Gls{dcg} is the weighted sum of the degree of relevance of the ranked items.
The weight is a decreasing function of the rank (position) of the object, and is therefore called discount.
In practice, the logarithmic decade with the rank $m$ has shown promising results. 
\ndcg normalizes \dcg by \idcg, which is simply the \dcg measure of the best ranking result i.e. the true ranking induced by target $\bc_i$.

For the load balancing task we can interpret the achievable rate $c_{ij}[n]$ as the relevance score of \BS $j$ for user $i$.
The higher the offered rate, the better is the ``rank" of the \BS.
Similar to the ranking problem, we want to preserve the ranking among the \BSs induced by $\bc_i$. 
Conventionally, \dcg is defined without the sigmoid function.
However, in our experiments we observe that reducing the dynamic range of $\bc_i$ using the sigmoid function helps train a better forecaster.
Both the \nmse and \ndcg are upper-bounded by $1$ and contribute equally to the loss. 


Computing \ndcg requires a sorting operation. 
Thus, the loss function \eqref{eq:joint_loss} is discontinuous---hence non-differentiable---and not amenable to optimization.
We instead use the smooth surrogate loss function proposed in \cite{pobrotyn2021neuralndcg} which has the following properties. 
First, it introduces a temperature parameter to approximate the sorting operation.
Thus, it makes it infinitely differentiable with non-sparse gradients. 
Second, it approaches the true \ndcg loss as the temperature parameter approaches zero.
In our experiments, we observe that using the smooth surrogate loss greatly improves the quality of the learned forecaster as compared to the indicator function.

\section{Solving the Control Problem using the Frank-Wolfe Method}
\label{sec:frank_wolfe_algo}
Even though the user association \mpc in \eqref{opt:util_max_mpc} is a convex program, solving it using a generic convex solver can be slow, especially for large cellular networks. 
Thus, we now present a computationally efficient algorithm using the Frank-Wolfe method \cite{jaggi13revisiting}, also known as the conditional gradient method, that is a low-memory, projection-free, first-order optimization method used for non-linear constrained optimizations.
While competing methods such as the projected gradient descent need a projection step into the constraint set, the Frank-Wolfe method instead solves a linear approximation---the first-order Taylor expansion---in the same set, thus always staying in the feasible set. 
In our case, we can analytically solve this linear optimization, making the Frank-Wolfe method a good fit for the presented user association formulation.

Consider the optimization problem in \eqref{opt:fw_generic} where $f(\cdot)$ is a differentiable function and $\cC$ represents a convex compact set.
\begin{subequations}
\label{opt:fw_generic}
\begin{align}
    \underset{\bz}{\mathrm{maximize}} &~f(\bz) \\
    \st~&\bz \in \cC.
\end{align}
\end{subequations}
Such constrained optimization problems can be solved iteratively using the Frank-Wolfe method.
In each iteration there are two key steps: the direction-finding step, and the update step.
In each iteration $k$th, we consider the linear approximation of the problem given by the first-order Taylor approximation of $f(\cdot)$ at the current solution $\bz_k$, and move in the direction the maximizes this linear approximation.
Specifically, if $\bs_k$ represents the solution of the following linear program
\begin{subequations}
\label{opt:fw_linear_oracle_general}
\begin{align}
    \underset{\bs}{\mathrm{maximize}}~&\bs^T\nabla f(\bz_k) \\
   \st~&\bs\in\cC,
\end{align}
\end{subequations}
then the update direction for the $k$th iteration is $\bd_k = \bs_k - \bz_k$.
Given the update direction, 
\begin{align}
    \bz_{k+1} & = \bz_k + \alpha \bd_k
\end{align}
where $\alpha$ represents the update step-size.
The algorithm terminates if the Frank-Wolfe gap $h_k = \bd^{T}_{k}\nabla f(\bz_k)$ falls below some tolerance.

For the proposed \mpc, let $f(\{x_{ij}[n]\})$ denote the objective in \eqref{opt:util_max_mpc_obj}.
Specifically,
\begin{align}
    \label{eq:fw_obj}
    f(\{x_{ij}[n]\}) &=  \sum_{n=1}^{H}\sum_{j\in\cB}\sum_{i\in\cU} x_{ij}[n] \log\left(\frac{\bar{c}_{ij}[n]}{\sum_{u\in\cU}x_{uj}[n]}\right)  + \frac{\eta}{2}\left(x_{ij}[n] - x_{ij}[n-1]\right)^2,
\end{align}
where $\bar{c}_{ij}[n]$ is as defined in \eqref{opt:util_max_mpc_rate_def} and the partial derivative of $f(\cdot)$ with respect to $x_{ij}[n]$ is given by
\begin{align}
    \label{eq:fw_partial_derivative}
    \frac{\partial f}{\partial x_{ij}[n]} =& \log(\bar{c}_{ij}[n]) - 1 - \log\left(\sum_{u\in\cU}x_{uj}[n]\right) \nonumber\\
    & + \eta(x_{ij}[n] - x_{ij}[n-1]) - \eta(x_{ij}[n+1] - x_{ij}[n])\mathbbm{1}(n<H).
\end{align}
Note that we use braces within the parenthesis to emphasize that $f(\cdot)$ is a multivariate function.
Then, at the $k$th iteration, the direction-finding linear program in \eqref{opt:fw_linear_oracle_general} can be written as,
\begin{subequations}
\label{opt:fw_linear_oracle}
\begin{align}
    \{s^{(k)}_{ij}[n]\} =& \underset{\{s_{ij}[n]\}}{\arg\max} \sum_{n=1}^H\sum_{j\in\cB}\sum_{i\in\cU} s_{ij}[n]\frac{\partial f}{\partial x^{(k)}_{ij}[n]} \\
   \st\nonumber\\
   &\sum_{j\in\cB}s_{ij}[n] = 1 \quad\forall i\in\cU, n=1, \dots, H\\
    & s_{ij}[n] \in [0, 1] \quad\forall i\in \cU, j\in\cB, n=1, \dots, H,
\end{align}
\end{subequations}
where $\{s^{(k)}_{ij}[n]\}$ represents the solution that maximizes the first-order Taylor approximation of $f(\cdot)$ at $\{x^{(k)}_{ij}[n]\}$.
The solution to the above constrained linear optimization is straight forward.
\begin{align}
    \label{eq:fw_linear_oracle_sol}
    j^*_{i}[n] &= \underset{j}{\arg\max} \frac{\partial f}{\partial x^{(k)}_{ij}[n]} \\
    s^{(k)}_{ij}[n] &= \mathbbm{1}(j = j^*_{i}[n]).
\end{align}
In other words, user $i$ chooses the \BS $j^*$ which maximizes the partial derivative at the $k$-th iteration.
Given $s^{(k)}_{ij}[n]$, the update direction is defined as $d^{(k)}_{ij}[n] = s^{(k)}_{ij}[n] - x^{(k)}_{ij}[n]$ and update step is given by
\begin{align}
    \label{eq:fw_update}
    x^{(k+1)}_{ij}[n] &= x^{(k)}_{ij}[n] + \alpha^{(k)}d^{(k)}_{ij}[n], 
\end{align}
where $\alpha^{(k)}$ is the step-size chosen according to the backtracking line-search rule \cite{pedregosa2020linearly}.
The pseudo-code for our Frank-Wolfe implementation is given in \algoref{algo:fw_btls}.
The function $Q(\cdot)$ in line 10 of \algoref{algo:fw_btls} is a quadratic upper bound on the curvature of $f(\cdot)$ and is defined as
\begin{align*}
    Q(\alpha, L) &= f(\{x_{ij}[n]\}) + \alpha h^{(k)} - \frac{\alpha^2L}{2}\sum^H_{n=1}\sum_{j\in\cB}\sum_{i\in\cU}d^{(k)}_{ij}[n],
\end{align*}
where $h^{(k)}$ (line 6) denotes the Frank-Wolfe gap at iteration $k$ and $L$ is the Lipschitz constant as defined in \cite{pedregosa2020linearly}.

\begin{algorithm}
\caption{Frank-Wolfe algorithm for user association \mpc ($\beta=0.9$, $\delta=2.0$)}\label{algo:fw_btls}
\begin{algorithmic}[1]
\Require $\bar{c}_{ij}[n],~x_{ij}[0]$
\Ensure $x_{ij}[n], n=1,2, \dots, H$
\State Randomly initialize $x^{(0)}_{ij}[n]$ such that $\sum_{i}x^{(0)}_{ij}[n] = 1$
\For{$k=1, 2, \dots$}
\State $j^*_i[n] = \arg\max_j~\partial f(x^{(k)})/\partial x_{ij}[n]$
\State $s^{(k)}_{ij}[n] = \mathbbm{1}(j = j^*_i[n])$ 
\State $d^{(k)}_{ij}[n] = s^{(k)}_{ij}[n] - x^{(k)}_{ij}[n]$
\State $h^{(k)} = \sum_{n=1}^H\sum_{j\in\cB}\sum_{i\in\cU} d^{(k)}_{ij}[n]\cdot\partial f(x^{(k)})/\partial x_{ij}[n]$ \Comment{Known as the Frank-Wolfe gap.}
\State // BTLS begins
\State $h^{(k)} = \beta h^{(k-1)}$
\State $\alpha^{(k)} = \min\left\{1, h^{(k)}/\left(h^{(k)}\sum_{n=1}^H\sum_{j\in\cB}\sum_{i\in\cU}\left(d^{(k)}_{ij}[n]\right)^2\right)\right\}$
\For {$f(\{x_{ij}^{(k)}[n] + \alpha^{(k)} d_{ij}^{(k)}[n]\}) < Q(\alpha^{(k)}, h^{(k)})$}
\State $h^{(k)} = \delta h^{(k)}$
\EndFor
\State //BTLS ends
\State $x^{(k+1)}_{ij}[n] = x^{(k)}_{ij}[n] + \alpha^k\cdot d^{(k)}_{ij}[n]$
\EndFor
\State\Return $x^{(k)}_{ij}[n]$
\end{algorithmic}
\end{algorithm}


\section{Simulation Results}
\label{sec:results}

\begin{table}
\caption{Wireless simulation parameters}
\label{tab:sim_param}
\centering
\begin{tabular}{|c|c|}
\hline
Transmit Power           & $46$ dBm                                                                     \\ \hline
Frequency Bands          & [$3750$ MHz, $750$ MHz]                                                        \\ \hline
Bandwidth                & \begin{tabular}[c]{@{}c@{}}$3700$ MHz: $40$ MHz\\ $700$ MHz: $10$ MHz\end{tabular} \\ \hline
Pathloss and shadowing & Urban macrocell model from 3GPP \cite{3GPP2017} \\ \hline 
Number of \BS                & $19\times 3\times 2 = 114$                                                                    \\ \hline
Number of users                & $1200$                                                              \\ \hline
User distribution               & Uniform or GMM with $10$ cluster centers                                                                \\ \hline
Inter-site separation              & $500$ m                                                                 \\ \hline
Noise                    & $-174$ + $10$ dBm/Hz                                                           \\ \hline
BS Height                & $25$ m                                                                       \\ \hline
User Height                & $1.5$ m                                                                      \\ \hline
\end{tabular}
\end{table}
We now evaluate the forecaster-aided user association strategy in \eqref{opt:util_max_mpc} through Monte-Carlo simulations.
The proposed approach is compared against a \maxcap association scheme, a genie-aided \mpc, and a \hit agnostic scheme, in terms of user service rates and number of \hos per time slots.
These baselines are described in detail in \secref{subsec:service_rate_ho}.
In \secref{subsec:mpc_vs_rl}, we compare the proposed forecaster-aided \mpc approach against model-free \rl from \cite{gupta21load} in terms of training sample-efficiency and generalization capabilities.
Finally, \secref{subsec:fw_vs_cvxpy} compares the execution time of our Frank-Wolfe implementation with a generic convex solver from CVXPY, and the \maxcap scheme.

\subsection{Evaluation Environment}
The wireless environment for the simulations was developed by Meta using their industry-grade system simulator.
We consider a $19$ site hexagonal deployment with an inter-site distance of $500$ m and $1200$ active users.
Each site supports three sectors and two frequency bands. 
Therefore, the total number of \BSs is $|\cB| = 19\times3\times2 = 114$.
For each Monte-Carlo iteration, $1200$ users are dropped either uniformly or according to a Gaussian mixture model (GMM) with $10$ cluster centers and diagonal covariance matrix, to simulate user clustering and inhomogeneous traffic conditions.
For GMM, during each Monte-Carlo iteration, the cluster centers are chosen uniformly, and the standard deviation in each dimension is $150$ m.
Users move with random speed $v$ drawn from a normal distribution with mean $15$ and variance $9$ \cite{3GPP_RRC_2020}, and their trajectory is simulated by a Gauss-Markov process \cite{tabassum19fundamentals}.
The simulator generates spatially consistent channel coefficients using statistical ray-tracing, which is crucial for a reliable mobility simulation.
It uses the urban macrocell pathloss model adopted by 3GPP \cite[page 26]{3GPP2017}, which incorporates the effects of blockage, the multi-slope nature of pathloss, and the elevation difference between the BSs and the UEs.
The \hit $\tho{i}$ is modeled as a discrete random variable. 
If a \ho is initiated, it is successful with a probability of $0.8$ and suffers a \hit of $20$ ms.
Otherwise, with a probability of $0.2$, it results in a link failure and suffers a \hit of $90.76$ ms. 
The total slot duration is $T_s = 100$ ms.
The link failure probability and interruption times were obtained through a network simulator-3 (ns-3) simulation\footnote{https://github.com/CollinBrady1993/ns-3-dev-git/tree/Handover} and the parameters for wireless simulation are listed in Table \ref{tab:sim_param}.

\subsection{Service Rates and Handovers}
\label{subsec:service_rate_ho}
\begin{figure}[t!]
    \centering
    \subfloat[Uniform user drop]{%
        \includegraphics[width=3in,keepaspectratio]
        {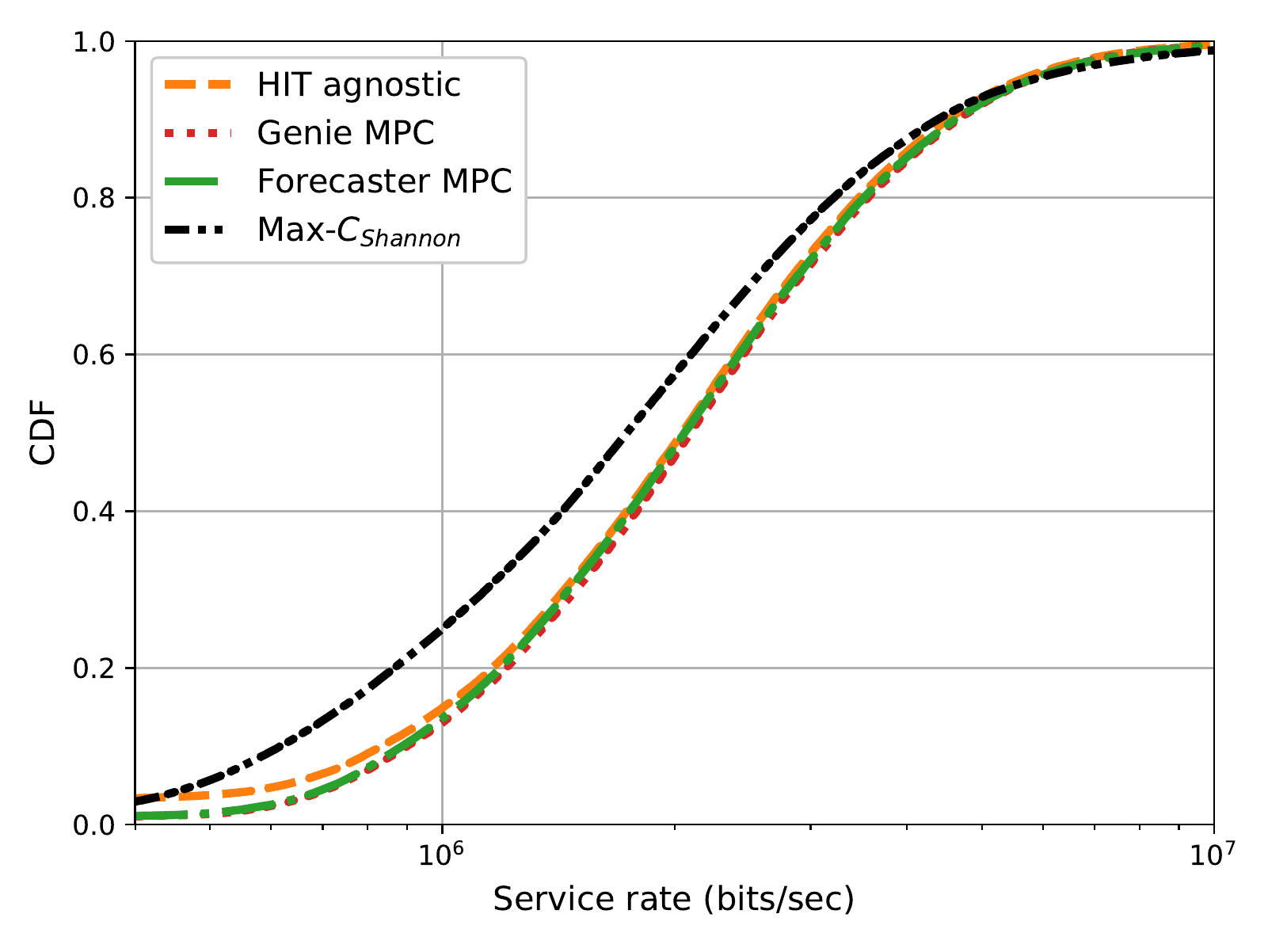}
        \label{fig:rate_cdf_uniform}}
    \quad
    \subfloat[Clustered user drop with $10$ centers]{%
        \includegraphics[width=3in,keepaspectratio]
        {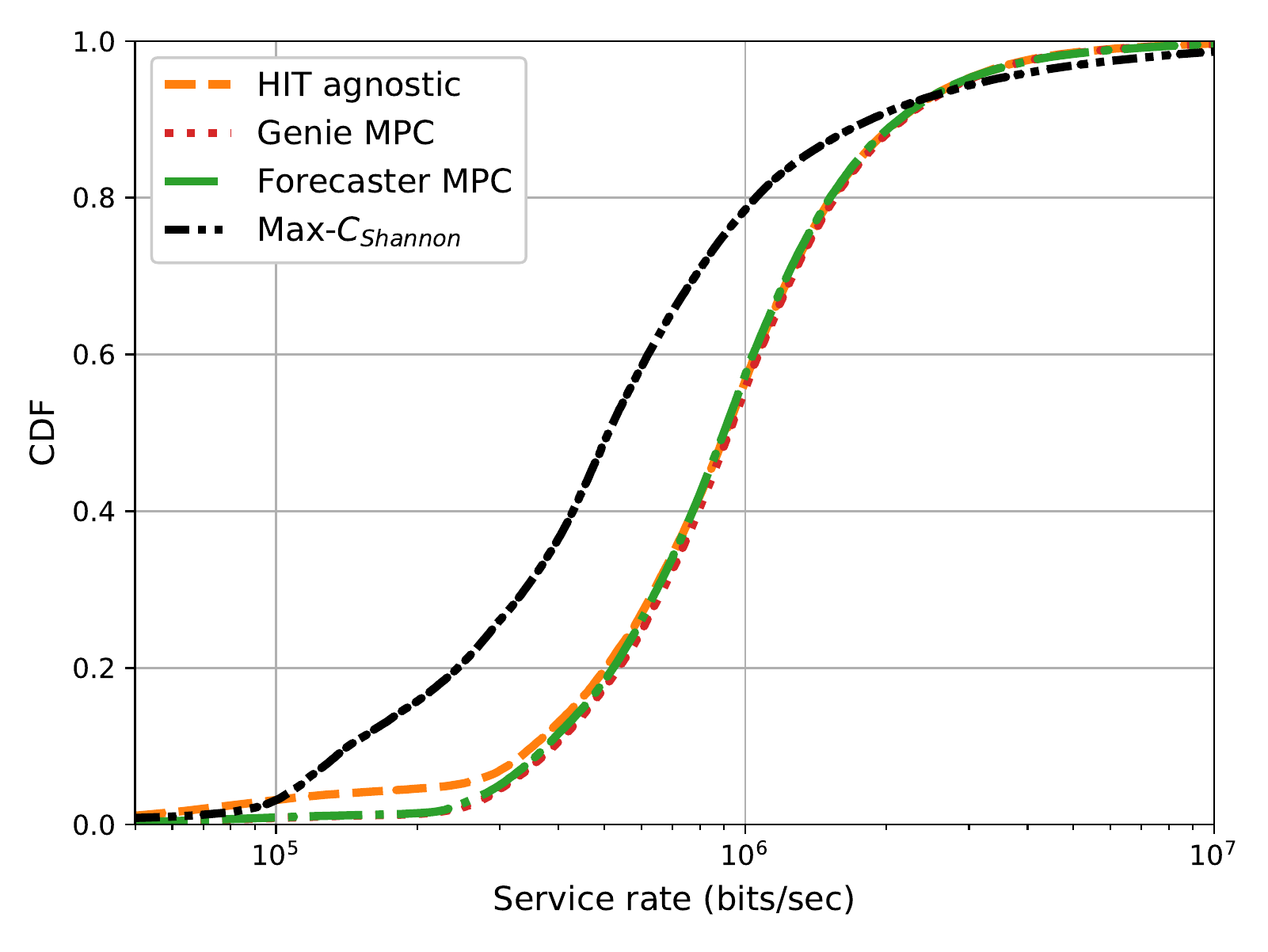}%
        \label{fig:rate_cdf_clustered}}
    \caption{Service rate CDF for the (a) uniform user drop and (b) clustered user drop. The load balancing gain over the \maxcap scheme is evident at the lower percentiles and more profound for an inhomogeneous traffic scenario.}
    \label{fig:rate_cdf}
\end{figure}

We will now compare the performance of the proposed forecaster-aided \mpc solution \eqref{opt:util_max_mpc} with the \maxcap user association, a \hit agnostic load balancing scheme, and to a genie-aided \mpc in terms of the user  service rate and the number of \hos triggered in a time slot.
Under the \maxcap scheme, the parent \BS assigns the user to the best \BS in terms of the achievable rate $c_{ij}[n]$.
For the \hit agnostic scheme, the central controller solves the utility maximization in \eqref{opt:util_max_mpc} for $H=1$ and $\eta=0$.
In other words, it only uses the current measurements (mobility-unaware) and does not penalize \hos.
This is similar to the user association strategies proposed in \cite{ye13user, shen14distributed}.
The genie-aided \mpc solves for $x_{ij}[n]$ using \eqref{opt:util_max_mpc} but assumes non-causal knowledge of future rates.

\figref{fig:rate_cdf} shows the service rate \cdf for the four user association schemes under different user drop scenarios where
\figref{fig:rate_cdf_uniform} presents uniform user drop and \figref{fig:rate_cdf_clustered} presents clustered user drop which emulates inhomogeneous traffic conditions.
The genie \mpc and the forecaster \mpc are evaluated for a forecasting horizon of $H=3$.
Compared to the \maxcap scheme, the log-utility-based association schemes provide a more uniform user experience by judiciously distributing users across \BSs.
Under the log-utility, the users with a strong wireless channel---for example, those close to the cellular tower---are offloaded to a \BS or frequency band with a relatively poor rate, leaving the lower frequency band for the users at the cell edge so that they can benefit from the lower pathloss and better penetration.
In \figref{fig:rate_cdf}, this results in significantly higher service rates at the 5th percentile compared to \maxcap, with a slight degradation beyond the 95th percentile.
We will refer to the improvement in service rate compared to \maxcap as \textit{rate gain}.
In \figref{fig:rate_cdf_uniform} and \figref{fig:rate_cdf_clustered}, we observe that compared to the uniform drop scenario, the 5th percentile rate gain is higher for the clustered user drop where the user traffic is inhomogeneous.
Due to this additional spatial heterogeneity, the user service rates suffer greater degradation under the load-agnostic \maxcap association strategy where \BSs close to the user clusters end up serving most of the users in the network while others remain under-utilized.
As a result, load-aware association is more beneficial for the clustered user drop where the 5th percentile rate gain is about $3.5\times$ as opposed to $1.5\times$ for the uniform user drop.
\begin{figure}[t!]
    \centering
    \subfloat[Uniform user drop]{%
        \includegraphics[width=3in,keepaspectratio]
        {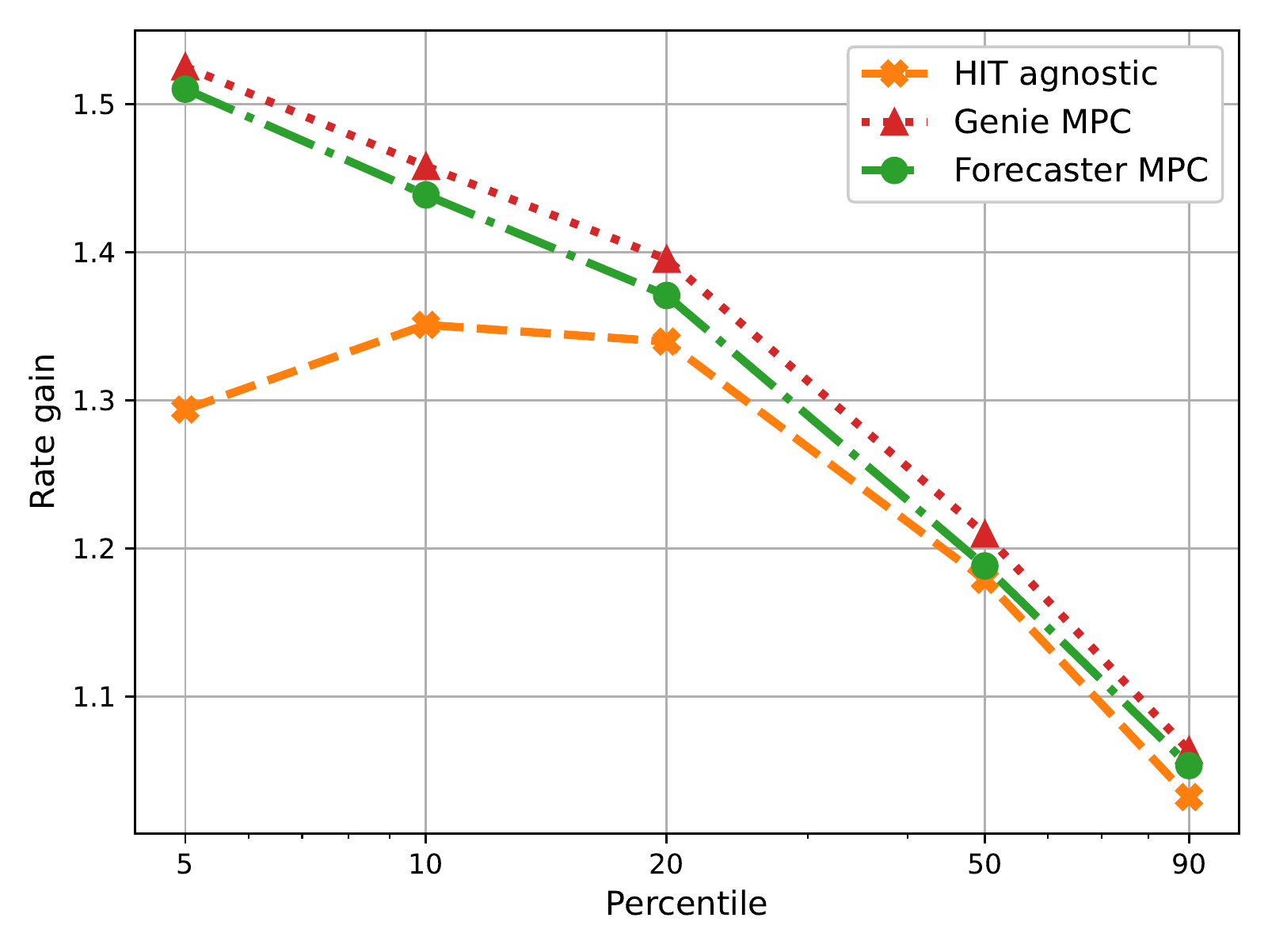}
        \label{fig:rate_percentile_uniform}}
    \quad
    \subfloat[Clustered user drop with $10$ centers]{%
        \includegraphics[width=3in,keepaspectratio]
        {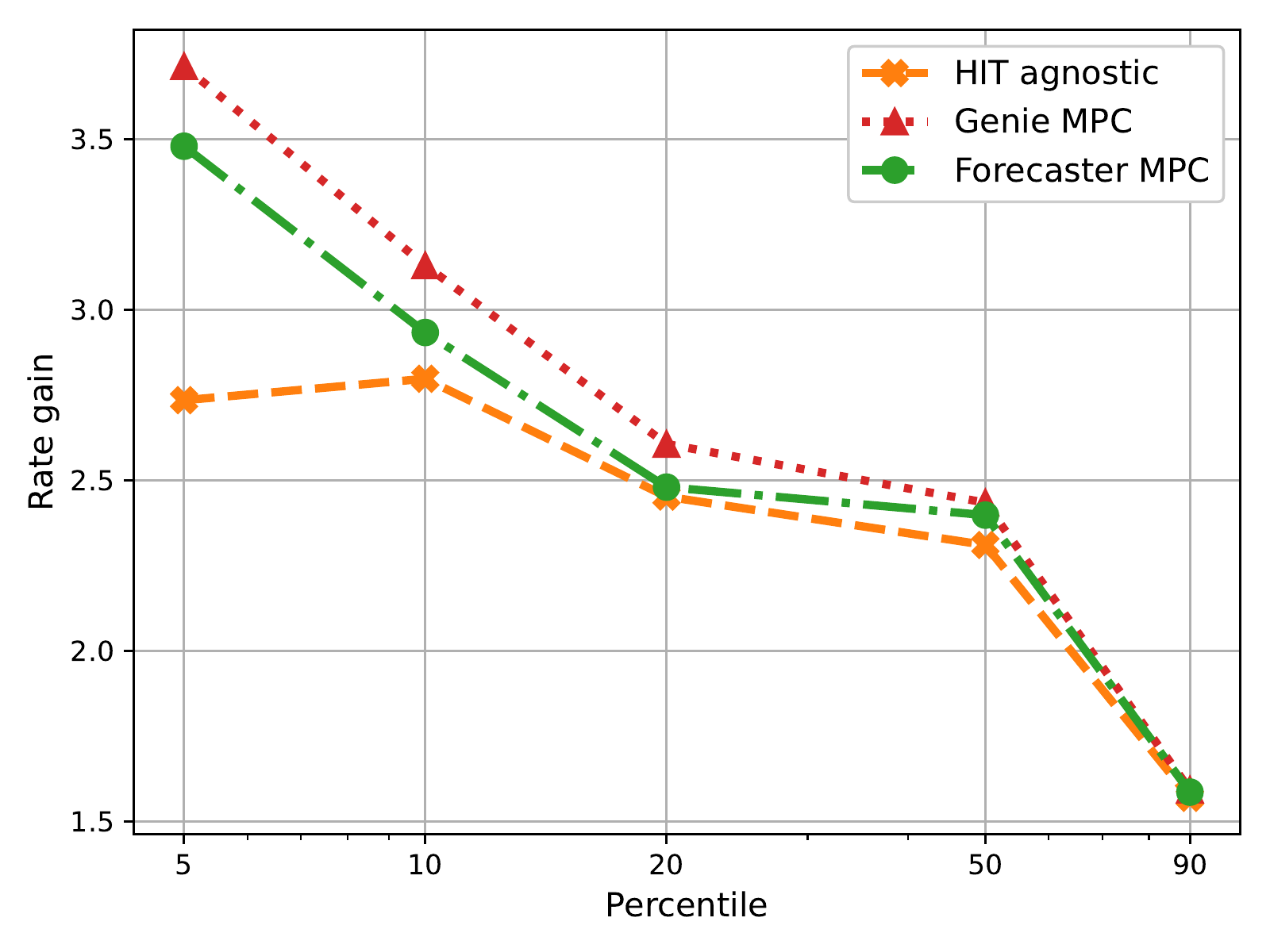}%
        \label{fig:rate_percentile_clustered}}
    \caption{Rate gain from load balancing across different percentiles for (a) uniform user drop and (b) clustered user drop. The forecaster \mpc achieves service rates similar to genie \mpc. The \hit agnostic approach suffers throughput degradation from \hos.}
    \label{fig:rate_percentile}
\end{figure}

The coverage improvement, as measured by the 5th percentile service rate, can be more clearly observed in \figref{fig:rate_percentile_uniform} and \figref{fig:rate_percentile_clustered} which present the rate gain from the three log-utility-based association schemes at different percentiles for both uniform and clustered user drops.
However, the gain diminishes at the higher percentiles because, compared to the \maxcap scheme, under the log-utility `good' users suffer a slight degradation in \sinr to help improve the overall network performance. 
Furthermore, \figref{fig:rate_percentile} shows that the forecaster \mpc achieves service rates similar to the genie \mpc, especially for the uniform user drop scenario.
Among the three load-aware association strategies, the mobility-aware \mpc approaches benefit from knowledge of future user trajectories and thus, can avoid \ho related degradation to users' service rates.
On the other hand, the \hit agnostic scheme is plagued with excessive \hos and as a result the edge user performance suffers.
In \figref{fig:rate_percentile_clustered}, the \hit agnostic scheme achieves a 5th percentile rate gain of about $2.6\times$ over \maxcap.
This means, compared to the proposed forecaster-aided \mpc, \hit agnostic scheme suffers a $39\%$ loss in edge user service rate due to additional \hos.

\begin{figure}[t!]
    \centering
    \subfloat[Uniform user drop]{%
        \includegraphics[width=3in,keepaspectratio]
        {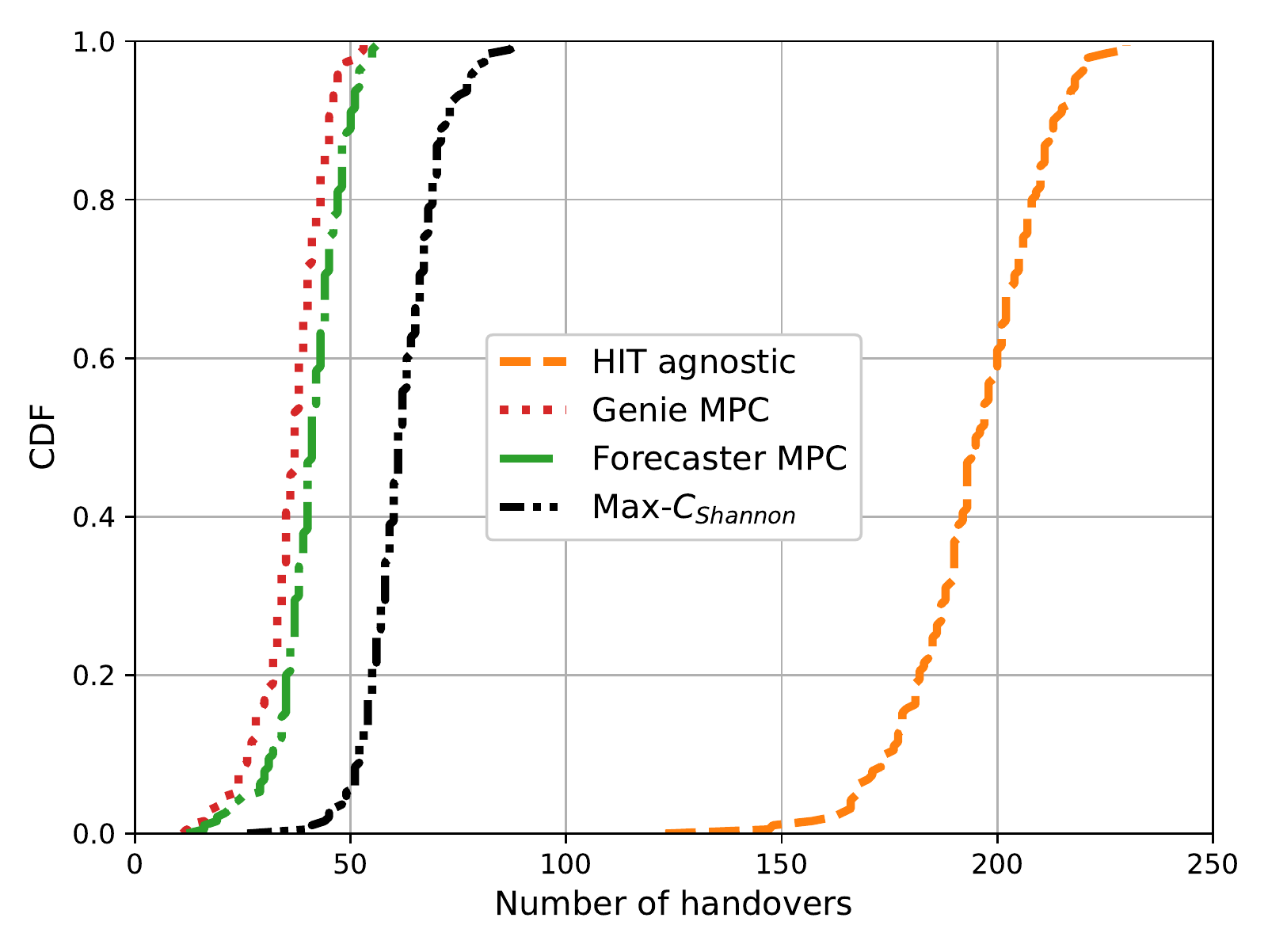}
        \label{fig:ho_cdf_uniform}}
    \quad
    \subfloat[Clustered user drop with $10$ centers]{%
        \includegraphics[width=3in,keepaspectratio]
        {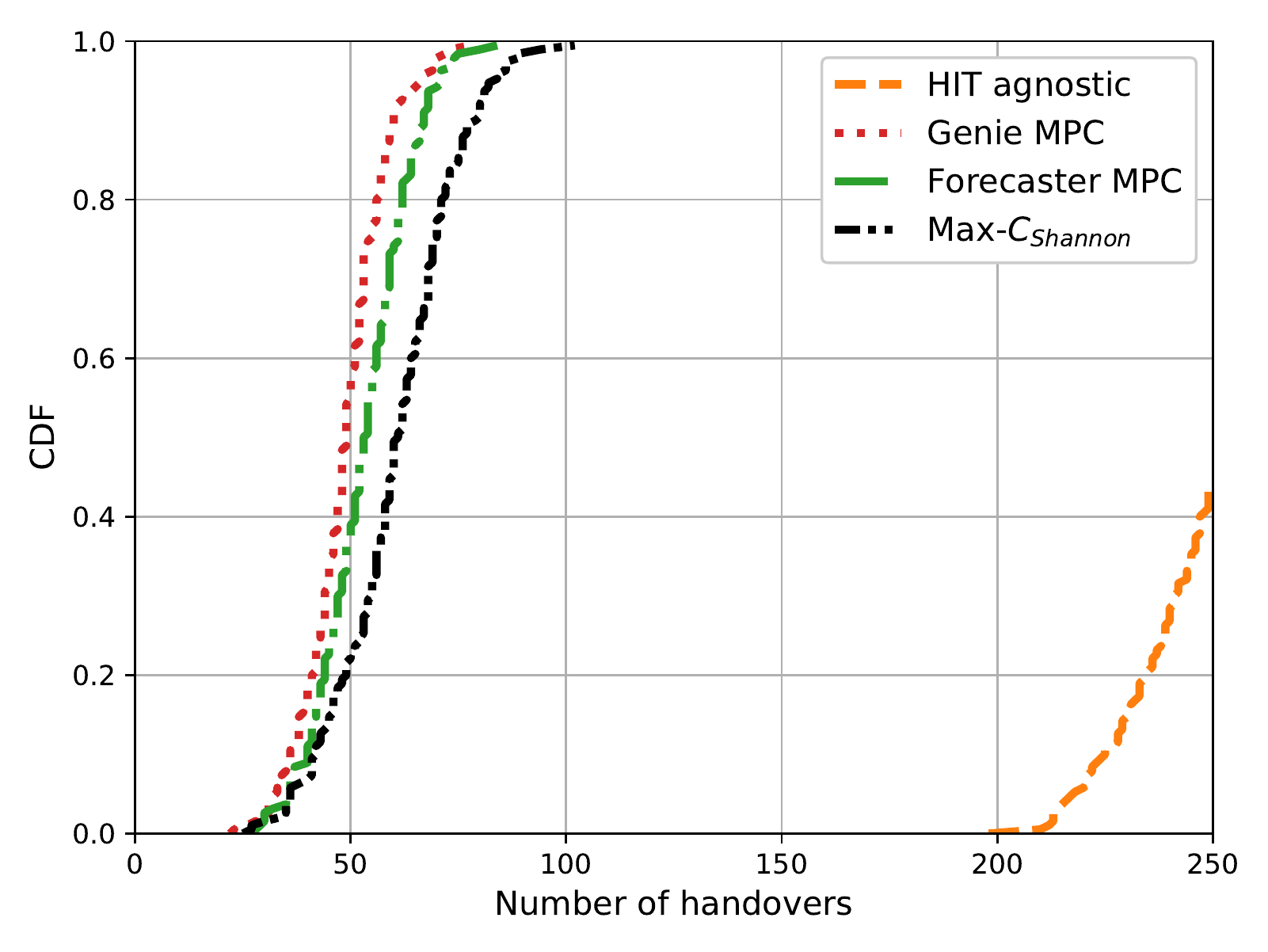}
        \label{fig:ho_cdf_clustered}}
    \caption{\cdf of \hos triggered in a time slot for (a) uniform user drop and (b) clustered user drop. Forecaster \mpc significantly reduces \hos compared to \hit agnostic by leveraging the available forecast of user rates.}
    \label{fig:ho_cdf}
\end{figure}
The \ho performance of the four association schemes is compared in \figref{fig:ho_cdf} where
\figref{fig:ho_cdf_uniform} and \figref{fig:ho_cdf_clustered} present the \cdf of the number of \hos initiated in a time slot. 
The \hit agnostic scheme initiates the most number of \hos, as evident from the \cdf, and the proposed forecaster-aided \mpc scheme reduces the median number of \hos by $7\times$ for the clustered user drop and by $5\times$ for the uniform user drop.
Compared to the genie-aided scheme, the median of \ho \cdf for the proposed scheme is about $1.5\times$ greater. 
Note, although not considered in this work, reducing \hos has benefits beyond improving user service rate. 
For example, unnecessary \hos and the associated signaling overhead can lead to increased session latency.
Similarly, reducing unwanted \hos will reduce energy consumption which is especially beneficial for mobile devices.
These results illustrate that using the proposed forecaster design with the novel loss function, the \mpc in \eqref{opt:util_max_mpc} can achieve performance close to the genie-aided scheme in terms of both the user  service rate and the number of \hos.

\subsection{\mpc vs \rl: Sample-efficiency \& Generalization}
\label{subsec:mpc_vs_rl}
\begin{figure}[t!]
    \centering
    \includegraphics[width=6in,                 keepaspectratio]{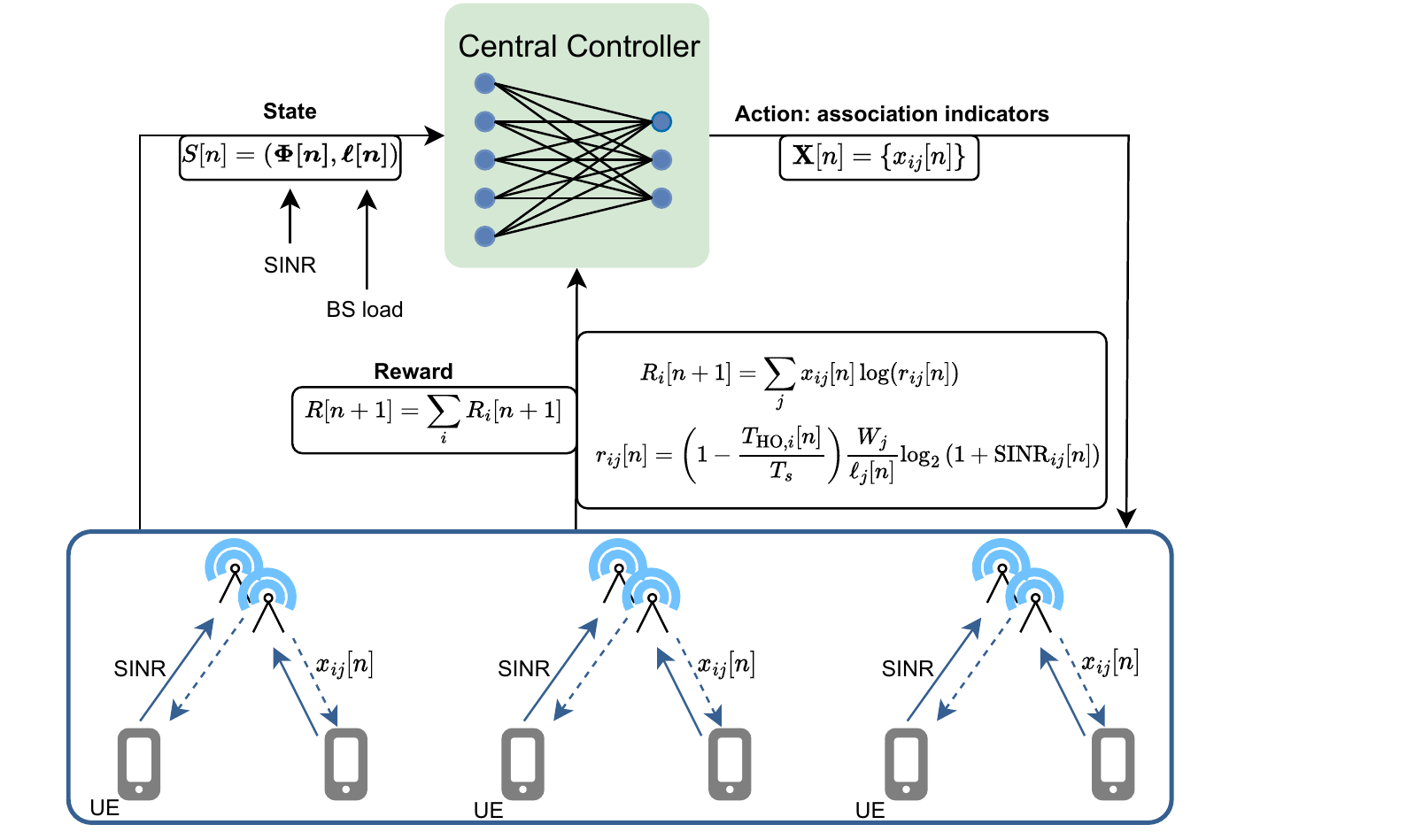}
    \caption{\mdp for mobility-aware user association.}
    \label{fig:mlb_mdp}
\end{figure}

In this section, we will compare the proposed \mpc approach to model-free \rl in terms of the training sample-efficiency and their ability to generalize to different user drops.
We use the \mdp model and the neural network architecture proposed in \cite{gupta21load}.
However, we consider a simulation environment with $10\times$ more users and \BSs.
As such, unlike our previous work \cite{gupta21load} we had little success with the deep recurrent Q-network (DRQN) algorithm and instead use the proximal policy optimization (PPO) algorithm \cite{schulman17proximal} for training.
The \mdp corresponding to mobility-aware user association  can be defined as the 5-tuple $\langle\cS,\cX,\cT,\cR,\gamma\rangle$. 
$\cS$ is the state space and $\cX$ represents the action space. 
At each time step $n$, the central controller receives a representation of the network's current \textit{state}, $S[n] \in \cS$, and on that basis executes an \textit{action}, $\bX[n] \in \cX$, causing the environment state to transition to a new state $S[n+1]$ with probability $\bbP(S[n+1]|S[n], \bX[n]) = \cT(S[n],\bX[n],S[n+1])$. 
In the next time step the central controller receives the \textit{reward} $R[n+1] = \cR(S[n], \bX[n], S[n+1])$ and the new state $S[n+1]$. 
Over an episode of $T$ time slots, the controller collects a cumulative reward $\sum_{n=1}^T\gamma^nR[n+1]$ where $\gamma \in (0, 1]$
is known as the discount factor introduced mainly to have a
finite cumulative reward, especially as $T$ tends to infinity. 
The objective of the controller is to maximize the expected cumulative reward. 
The specific formulations of the state, action, and reward are given below and summarized in \figref{fig:mlb_mdp}.

\textbf{State:} The central controller chooses the new association based on the $|\cU|\times|\cB|$ \sinr measurements $\bm{\Phi}[n]$ received for each user-\BS pair and the $|\cB|\times 1$ BS load $\bm{\ell}[n]$. 
Thus, the state at time $n$ is defined as a tuple
\begin{align}
    \label{eq:state_def}
    S[n] = (\bm{\Phi}[n], \bm{\ell}[n]).
\end{align}

\textbf{Action:} Observing the \sinr for user-\BS pairs and the load on each \BS, the central controller returns the new association indicators $\{x_{ij}[n]|i\in\cU, j\in\cB\}$ such that they satisfy \eqref{opt:util_max_genie_ue_cons} and \eqref{opt:util_max_genie_indicator_cons}. 
The action at time $n$ is thus denoted by a $|\cU|\times|\cB|$ binary matrix $\bX[n]$ such that the sum of each row is $1$.

\textbf{Reward:} Given the association matrix $\bX[n]$, the serving BS for user $i$ is given by $j^* = \arg\max_jx_{ij}[n]$ and it receives utility $\log(r_{ij^*}[n])$ where $r_{ij^*}[n]$ is given by \eqref{eq:service_rate_def}. 
The \hit $T_{{\rm HO},i}[n]$ is non-zero if user $i$ changes serving \BSs, $x_{ij^*}[n] \neq x_{ij^*}[n-1]$. 
A \ho can either be successful or can result in an link failure. 
If an link failure event occurs, the user re-establishes its connection with the network via the radio resource control (RRC) protocol, which results in a large interruption time of the order of hundreds of milliseconds. 
The reward received by the controller at time $n+1$ is the sum of utility received by all users.
\begin{align}
\label{eq:reward_def}
    R[n+1] &= \sum_{j\in\cB}\sum_{i\in \cU}x_{ij}[n] \log\left(r_{ij}[n]\right)
\end{align}
Note that $\sum_{n=1}^TR[n+1]$ is equal to the objective in \eqref{opt:util_max_genie_obj}.

\textbf{State transition probabilities:} The state $S[n]$ has two parts. 
The evolution of $\bm{\Phi}[n]$ is governed by the motion of users and is independent of actions taken by the controller. 
However, the load on each BS depends on the past action $\bX[n-1]$ and the results of the \ho procedure. 
Neither user mobility nor the \ho procedure are explicitly modeled and the \rl agent must learn the underlying system dynamics and determine the optimal policy in a model-free manner.
\begin{figure}
    \centering
    \includegraphics[width=4in,keepaspectratio]{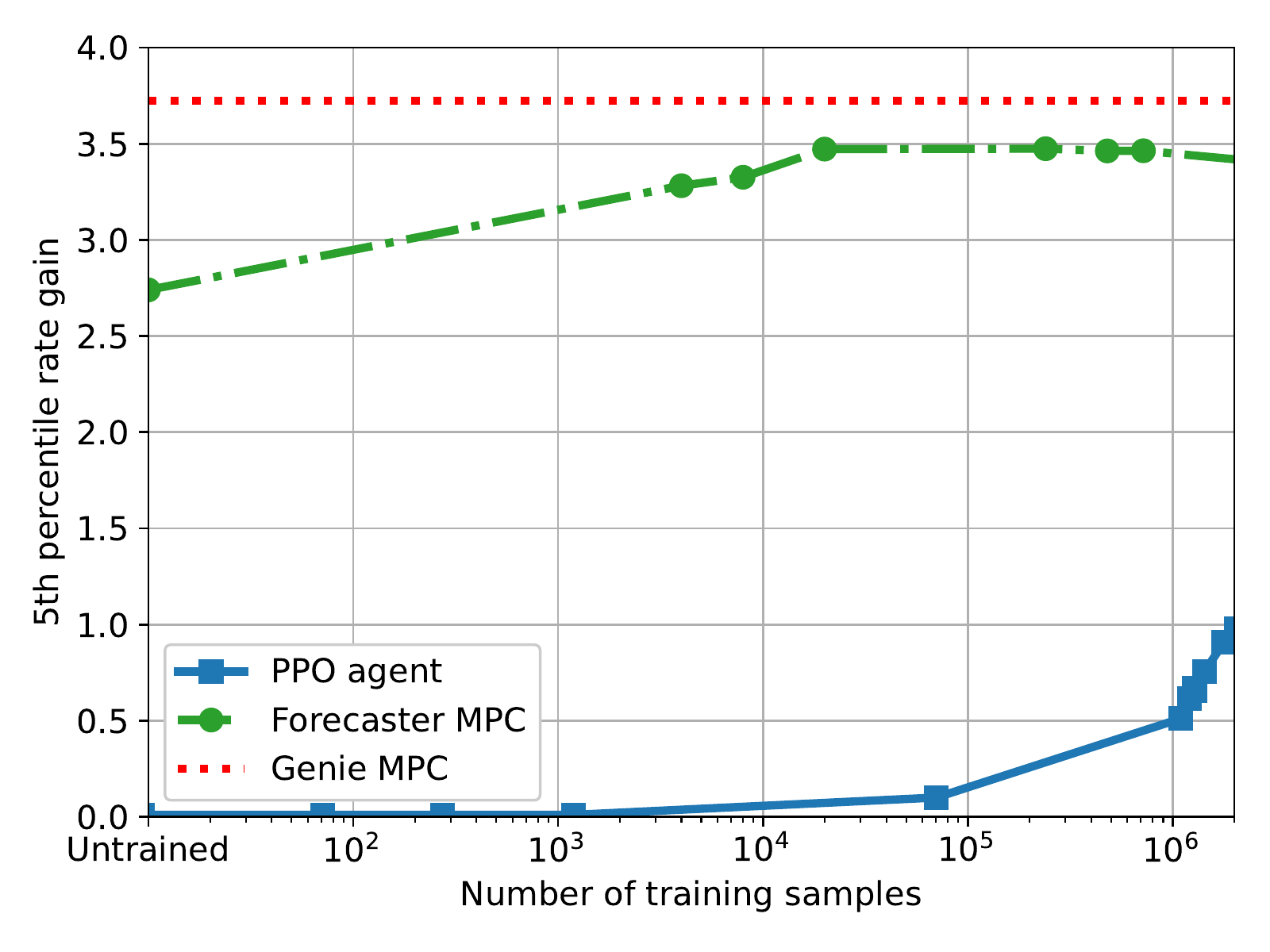}%
    \caption{Comparing training sample efficiency of user rate forecaster and \rl agent using the PPO algorithm. Even with $100\times$ more training samples, the PPO agent achieves 5th percentile service rate similar to \maxcap.}
    \label{fig:sample_efficiency}
\end{figure}

For this \mdp model, an \rl agent is trained for centralized control using the PPO algorithm. \figref{fig:sample_efficiency} presents the growth of 5th percentile rate gain of model-free \rl and the proposed forecaster \mpc with the number of training samples.
We train the PPO agent and the forecaster the same dataset.
A training sample for the PPO agent refers to the $4$-tuple $(S[n], \bX[n], R[n+1], S[n+1])$, as is typical for \rl. 
For the forecaster, a training sample refers to a labeled sample for supervised training as defined in \secref{sec:forecaster_design}, where the input feature is the time-series of current and past capacity measurements and the target label is the measurement for future time steps.
Note, that one \rl training sample includes measurements from all users in the network. 
However, the forecaster predicts the trajectory for each user independently and thus, treats measurements from each user as a separate training sample.
We observe in \figref{fig:sample_efficiency} that the forecaster needs about $10^4$ training samples to achieve performance close to the genie.
However, even with $100\times$ more training samples, the PPO agent can only achieve a edge user service rate similar to \maxcap, i.e. a 5th percentile rate gain of $1$.
In terms of training time, training the PPO agent with $10^6$ samples took about 3 weeks, where the forecaster can be trained in a few hours.
Both these models were trained on a machine with an Intel(R) Xeon(R) Gold 6226R CPU @ 2.90GHz and NVIDIA RTX A5000 GPU.
Note that in \figref{fig:sample_efficiency}, even an untrained forecaster can improve the edge user service rate by $2.7\times$.
This is because it has access to the true capacity measurements for the current time step.
Thus, even with random predictions from an untrained forecaster and reliable measurements for the current slot, the \mpc scheme leads to a well-balanced user distribution across bands and \BSs.
This is another advantage of our model-based approach over model-free \rl.
The gap between untrained forecaster-aided \mpc and the genie-aided \mpc represents the forecasting gain at $H=3$.

\begin{figure}
    \centering
    \subfloat[Uniform drop]{%
        \includegraphics[width=3.15in,
                        keepaspectratio]
        {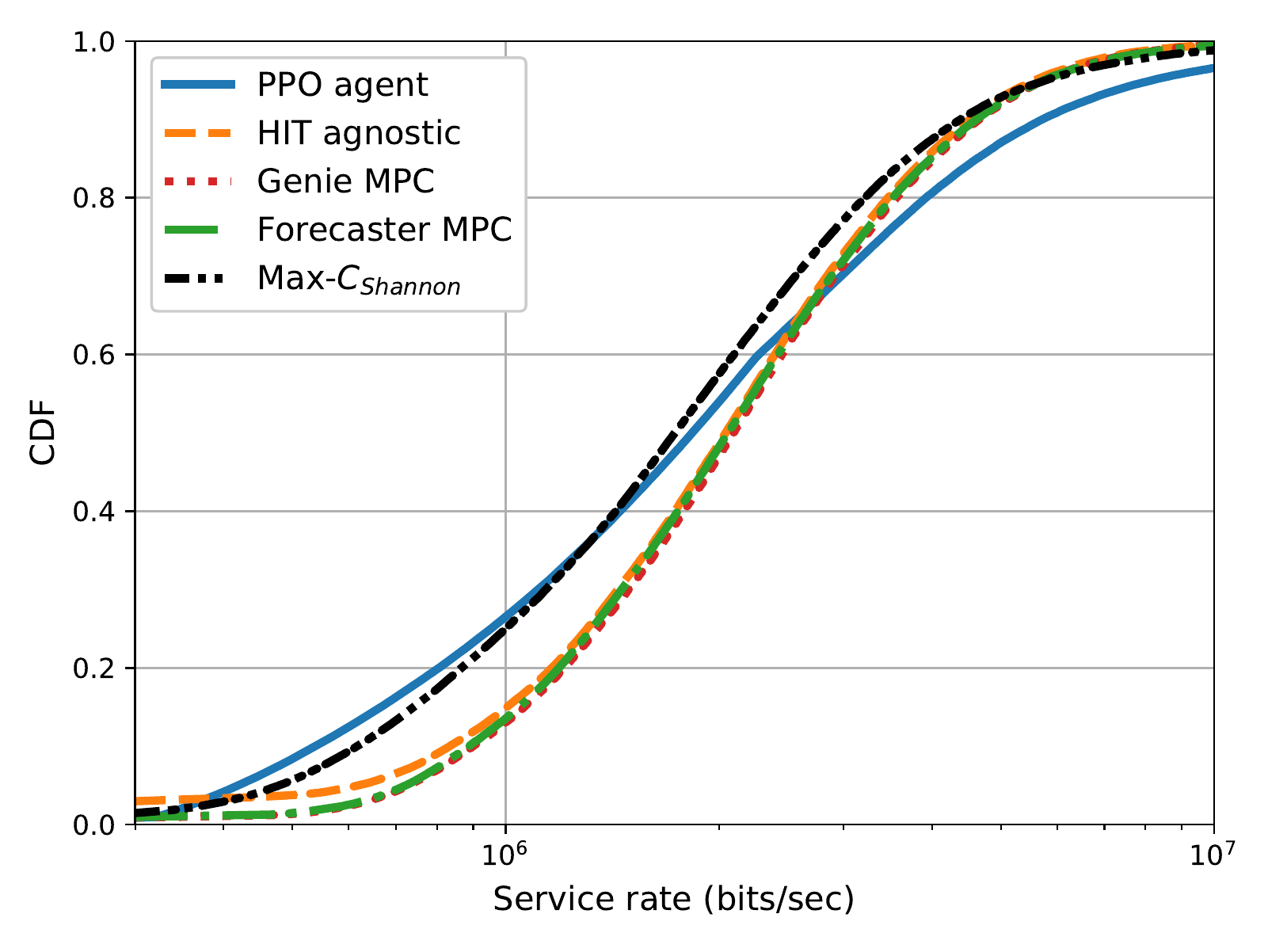}%
        \label{fig:ppo_rate_cdf_uniform}}
    \quad
    \subfloat[Clustered drop]{%
        \includegraphics[width=3.15in,
                        keepaspectratio]
        {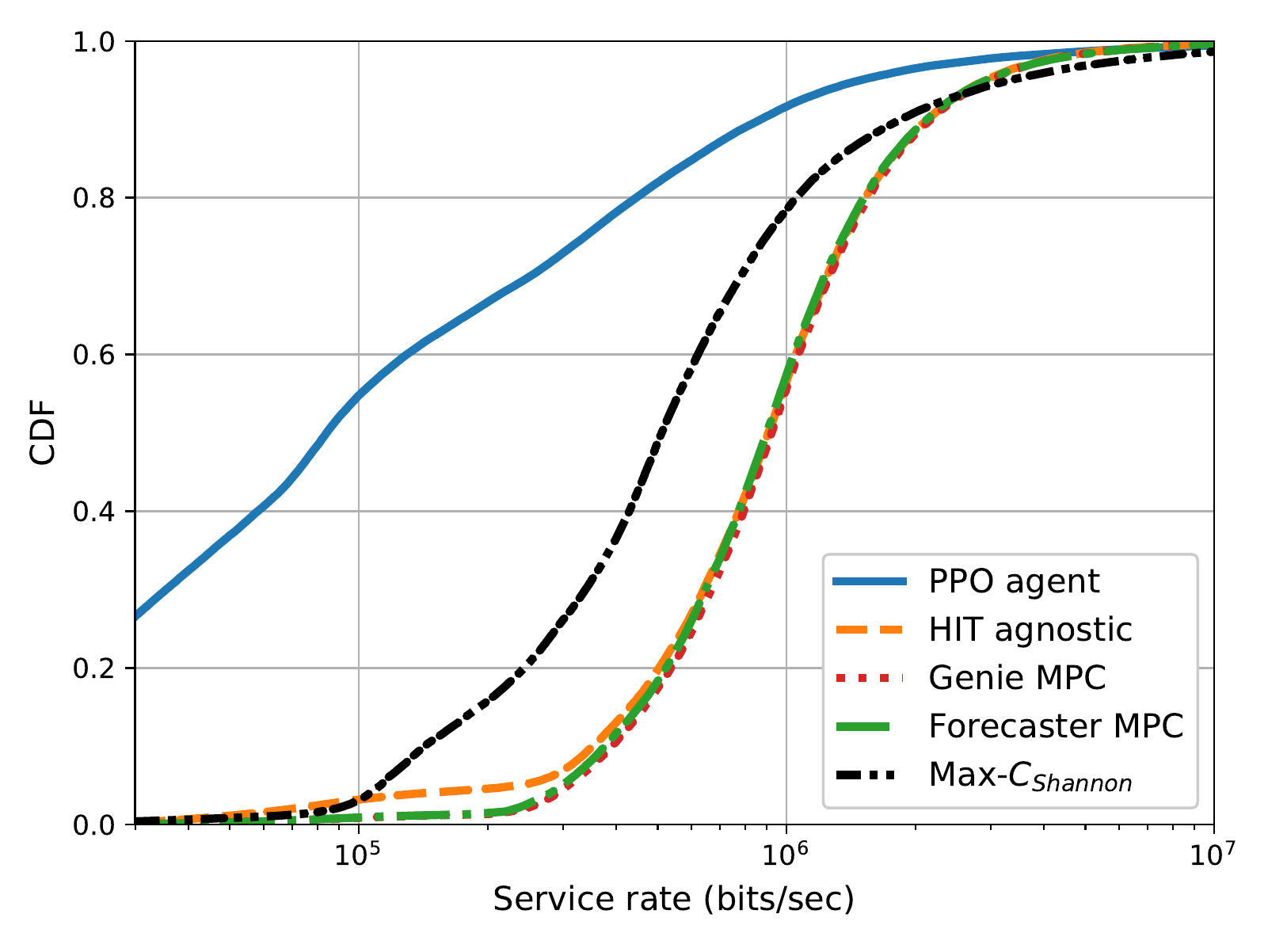}%
        \label{fig:ppo_rate_cdf_clustered}}
    \caption{Comparing the generalization capability of \rl and the rate forecaster. The PPO agent and the forecaster are trained on (a) uniform user drop scenario. These trained models are then used for inference on (b) a clustered user drop. PPO agent exhibits poor generalization while the forecaster reliably learns the user mobility patterns and generalizes across user drops.}
    \label{fig:ppo_rate_cdf}
\end{figure}

Next, we train the forecaster using $10^4$ training samples and the PPO agent using $10^6$ training samples from the uniform user drop scenario. 
\figref{fig:ppo_rate_cdf_uniform} shows the \cdf of the service rates achieved by the two user association strategies on the uniform user drop. 
We observe that the forecaster-aided \mpc achieves performance close to the genie and the PPO agent achieves performance similar to the \maxcap strategy.
These trained models are then tested on clustered user drops and the corresponding service rate \glspl{cdf} are presented in \figref{fig:ppo_rate_cdf_clustered}. 
While the forecaster \mpc can effectively translate its learned mobility model to the clustered drop scenarios, the PPO agent suffers significant degradation and exhibits poor generalization.
This is probably because the PPO agent learns some specific attribute of the uniform drop dataset which does not translate to the clustered drop scenario.
Such uninterpretability and poor generalization of \rl methods have been well documented and methods to combat such behavior are an active area of research.

\subsection{Computation Efficiency of Frank-Wolfe}
\label{subsec:fw_vs_cvxpy}
In the next section we evaluate the computational efficiency our optimization algorithm which is based on the Frank-Wolfe method.
We compare our Frank-Wolfe-based implementation to a splitting conic solver (SCS) implementation from CVXPY \cite{cvxpy1}, a standard software tool for convex optimization.
In \tabref{tab:obj_value} we present the user service rates, the cell sum-rates, and the log-utility achieved by the two optimization algorithms for different forecasting horizons.
The cell sum-rate for a \BS $j$ is defined as $\sum_i x_{ij}[n]r_{ij}[n]$.
The two algorithms achieve similar performance on all three metrics and this validates that our algorithm converges to the correct global optima. 

In \tabref{tab:compute_time}, we present the per time slot execution time for solving the proposed \mpc in \eqref{opt:util_max_mpc} for $H=1$ and $H=2$. 
The execution times in \tabref{tab:compute_time} are normalized by the execution time of the \maxcap scheme.
Note, the tables present execution time for solving the \mpc which does not include the interference time from the forecaster.
We observe that on our fairly large network with $114$ \BSs and $1200$ users, the SCS solver is quite computationally expensive.
However, our Frank-Wolfe implementation reduces the execution time by two orders of magnitude for a time horizon of $H=1$ and $H=2$.
This is because the direction-finding step in the Frank-Wolfe method \eqref{opt:fw_linear_oracle} poses a closed-form solution which was discussed in \secref{sec:frank_wolfe_algo}.
As such, the solution to the linear program reduces to a simple maximization over an array.
Using the backtracking line-search procedure proposed in \cite{pedregosa2020linearly} also aides in faster convergence by reducing the number of iterative steps.
These effects together make the proposed \mpc solution practically viable.

\begin{table}
\caption{Average service rate and objective achieved by splitting conic solver (SCS) from CVXPY and our optimization algorithm based on the Frank-Wolfe method.}
\label{tab:obj_value}
\centering
\begin{tabular}{|c|ccc|c|c|}
\hline
\multirow{2}{*}{} & \multicolumn{3}{c|}{User service rate (Mbps)}                                            & \multirow{2}{*}{\begin{tabular}[c]{@{}c@{}}Mean cell\\ sum-rate\end{tabular}}        & \multirow{2}{*}{\begin{tabular}[c]{@{}c@{}}Mean\\ objective value\end{tabular}} \\ \cline{2-4}
                  & \multicolumn{1}{c|}{5th percentile} & \multicolumn{1}{c|}{50th percentile} & 90th percentile &            &                                                                       \\ \hline
CVXPY + SCS       & \multicolumn{1}{c|}{0.69}        & \multicolumn{1}{c|}{2.06}        & 4.54       & 25.98 & 17410.72                                                                        \\ \hline
Our method      & \multicolumn{1}{c|}{0.70}        & \multicolumn{1}{c|}{2.04}        & 4.66       & 26.09 & 17411.60                                                                       \\ \hline
\end{tabular}
\end{table}

\begin{table}
\caption{Average per time slot execution time relative to the \maxcap scheme}
\label{tab:compute_time}
\centering
\begin{tabular}{|c|c|c|}
\hline
                      & \begin{tabular}[c]{@{}c@{}}Forecaster-aided  \\ \mpc for $H=1$\end{tabular} & \begin{tabular}[c]{@{}c@{}}Forecaster-aided  \\ \mpc for $H=2$\end{tabular}\\ \hline
CVXPY + SCS           & $137.5$          & $187.5$          \\ \hline
Our Method & $1.25$           & $3.26$           \\ \hline
\end{tabular}
\end{table}
\section{Conclusion}
\label{Sec:Conc}
We propose a learning-based \mpc policy for user association in multi-band heterogeneous networks with mobile users. 
Using the proposed novel loss function---linear combination of \nmse and \ndcg loss---the forecaster-aided approach is a sample-efficient and interpretable alternative to model-free \rl. 
We show that the service rates for the edge users can be greatly improved and our mobility-aware policy significantly reduces the \ho rate.
Moreover, using our implementation of the Frank-Wolfe method, the \mpc problem can be solved efficiently motivating the usefulness of our approach in a real deployment.
There is also considerable scope for further work. 
For example, designing a forecaster robust to out-of-distribution user mobility patterns, testing the performance on more sophisticated BS deployments with multiple tiers and more frequency bands, and a more rigorous study that models the \ho mechanism beyond the physical layer.
Extending the proposed techniques to MIMO systems and higher frequency bands (for example, millimeter wave and sub-terahertz) to study the impact of beamforming and high pathloss would also be suitable for the \mpc framework.

\section{Acknowledgements}
\label{Sec:Acks}
The authors would like to thank Ruichen Jiang at UT Austin for the insightful discussions on optimization algorithms which lead to the development of our computationally efficient optimization algorithm and for his very helpful feedback on the Frank-Wolfe method which significantly reduced the simulation time. 

\bibliographystyle{IEEEtran}
\bibliography{MananRef}

\begin{thebibliography}{10}
\providecommand{\url}[1]{#1}
\csname url@samestyle\endcsname
\providecommand{\newblock}{\relax}
\providecommand{\bibinfo}[2]{#2}
\providecommand{\BIBentrySTDinterwordspacing}{\spaceskip=0pt\relax}
\providecommand{\BIBentryALTinterwordstretchfactor}{4}
\providecommand{\BIBentryALTinterwordspacing}{\spaceskip=\fontdimen2\font plus
\BIBentryALTinterwordstretchfactor\fontdimen3\font minus
  \fontdimen4\font\relax}
\providecommand{\BIBforeignlanguage}[2]{{%
\expandafter\ifx\csname l@#1\endcsname\relax
\typeout{** WARNING: IEEEtran.bst: No hyphenation pattern has been}%
\typeout{** loaded for the language `#1'. Using the pattern for}%
\typeout{** the default language instead.}%
\else
\language=\csname l@#1\endcsname
\fi
#2}}
\providecommand{\BIBdecl}{\relax}
\BIBdecl

\bibitem{Seven_Andrews13}
J.~G. {Andrews}, ``Seven ways that {HetNets} are a cellular paradigm shift,''
  \emph{{IEEE} Commun. Mag.}, vol.~51, no.~3, pp. 136--144, Mar. 2013.

\bibitem{andrews14overview}
J.~G. Andrews, S.~Singh, Q.~Ye, X.~Lin, and H.~S. Dhillon, ``An overview of
  load balancing in {HetNets}: old myths and open problems,'' \emph{{IEEE}
  Wireless Commun.}, vol.~21, no.~2, pp. 18--25, Apr. 2014.

\bibitem{Arshad_Velocity17}
R.~{Arshad}, H.~{ElSawy}, S.~{Sorour}, T.~Y. {Al-Naffouri}, and M.~{Alouini},
  ``Velocity-aware handover management in two-tier cellular networks,''
  \emph{{IEEE} Trans. Wireless Commun.}, vol.~16, no.~3, pp. 1851--1867, Jan.
  2017.

\bibitem{feriani22multiobjective}
A.~Feriani \emph{et~al.}, ``Multiobjective load balancing for multiband
  downlink cellular networks: A meta-reinforcement learning approach,''
  \emph{{IEEE} J. Sel. Areas Commun.}, vol.~40, no.~9, pp. 2614--2629, Sept.
  2022.

\bibitem{singh13offloading}
S.~Singh, H.~S. Dhillon, and J.~G. Andrews, ``Offloading in heterogeneous
  networks: Modeling, analysis, and design insights,'' \emph{{IEEE} Trans.
  Wireless Commun.}, vol.~12, no.~5, pp. 2484--2497, May 2013.

\bibitem{kassir22analysis}
S.~Kassir, G.~de~Veciana, N.~Wang, X.~Wang, and P.~Palacharla, ``Analysis of
  opportunistic relaying and load balancing gains through v2v clustering,''
  \emph{{IEEE} Trans. Veh. Technol.}, vol.~71, no.~9, pp. 9896--9911, Sept.
  2022.

\bibitem{Modeling_Lin13}
X.~{Lin}, J.~G. {Andrews}, and A.~{Ghosh}, ``Modeling, analysis and design for
  carrier aggregation in heterogeneous cellular networks,'' \emph{{IEEE} Trans.
  Commun.}, vol.~61, no.~9, pp. 4002--4015, July 2013.

\bibitem{shen14distributed}
K.~Shen and W.~Yu, ``Distributed pricing-based user association for downlink
  heterogeneous cellular networks,'' \emph{{IEEE} J. Sel. Areas Commun.},
  vol.~32, no.~6, pp. 1100--1113, June 2014.

\bibitem{ye13user}
Q.~{Ye}, B.~{Rong}, Y.~{Chen}, M.~{Al-Shalash}, C.~{Caramanis}, and J.~G.
  {Andrews}, ``User association for load balancing in heterogeneous cellular
  networks,'' \emph{{IEEE} Trans. Wireless Commun.}, vol.~12, no.~6, pp.
  2706--2716, June 2013.

\bibitem{kim12distributed}
H.~Kim, G.~de~Veciana, X.~Yang, and M.~Venkatachalam, ``Distributed
  $\alpha$-optimal user association and cell load balancing in wireless
  networks,'' \emph{IEEE/ACM Trans. on Netw.}, vol.~20, no.~1, pp. 177--190,
  Feb. 2012.

\bibitem{ye16user}
Q.~Ye, O.~Y. Bursalioglu, H.~C. Papadopoulos, C.~Caramanis, and J.~G. Andrews,
  ``User association and interference management in massive mimo hetnets,''
  \emph{{IEEE} Trans. Commun.}, vol.~64, no.~5, pp. 2049--2065, May 2016.

\bibitem{liu19joint}
R.~Liu, Q.~Chen, G.~Yu, and G.~Y. Li, ``Joint user association and resource
  allocation for multi-band millimeter-wave heterogeneous networks,''
  \emph{{IEEE} Trans. Commun.}, vol.~67, no.~12, pp. 8502--8516, Dec. 2019.

\bibitem{teng21joint}
W.~Teng \emph{et~al.}, ``Joint optimization of base station activation and user
  association in ultra dense networks under traffic uncertainty,'' \emph{{IEEE}
  Trans. Commun.}, vol.~69, no.~9, pp. 6079--6092, Sept. 2021.

\bibitem{addali19dynamic}
K.~M. Addali, S.~Y. Bani~Melhem, Y.~Khamayseh, Z.~Zhang, and M.~Kadoch,
  ``Dynamic mobility load balancing for {5G} small-cell networks based on
  utility functions,'' \emph{IEEE Access}, vol.~7, pp. 126\,998--127\,011,
  Sept. 2019.

\bibitem{lin13towards}
X.~Lin, R.~K. Ganti, P.~J. Fleming, and J.~G. Andrews, ``Towards understanding
  the fundamentals of mobility in cellular networks,'' \emph{{IEEE} Trans.
  Wireless Commun.}, vol.~12, no.~4, pp. 1686--1698, Apr. 2013.

\bibitem{hsueh17equivalent}
S.-Y. Hsueh and K.-H. Liu, ``An equivalent analysis for handoff probability in
  heterogeneous cellular networks,'' \emph{{IEEE} Commun. Lett.}, vol.~21,
  no.~6, pp. 1405--1408, June 2017.

\bibitem{tabassum19fundamentals}
H.~Tabassum, M.~Salehi, and E.~Hossain, ``Fundamentals of mobility-aware
  performance characterization of cellular networks: A tutorial,'' \emph{{IEEE}
  Commun. Surv. Tutor.}, vol.~21, no.~3, pp. 2288--2308, thirdquarter 2019.

\bibitem{rakotomanana16optimum}
E.~Rakotomanana and F.~Gagnon, ``Optimum biasing for cell load balancing under
  {QoS} and interference management in {HetNets},'' \emph{IEEE Access}, vol.~4,
  pp. 5196--5208, Sept. 2016.

\bibitem{ghosh12heterogeneous}
A.~Ghosh \emph{et~al.}, ``Heterogeneous cellular networks: From theory to
  practice,'' \emph{{IEEE} Commun. Mag.}, vol.~50, no.~6, pp. 54--64, June
  2012.

\bibitem{QualcommHeNets}
A.~Damnjanovic \emph{et~al.}, ``A survey on {3GPP} heterogeneous networks,''
  \emph{{IEEE} Wireless Commun.}, vol.~18, no.~3, pp. 10--21, June 2011.

\bibitem{lin22embracing}
\BIBentryALTinterwordspacing
X.~Lin, L.~Kundu, C.~Dick, and S.~Velayutham, ``Embracing {AI} in 5{G}-advanced
  towards {6G}: A joint {3GPP} and {O-RAN} perspective,'' \emph{CoRR}, vol.
  abs/2209.04987, Sept. 2022. [Online]. Available:
  \url{https://arxiv.org/abs/2209.04987}
\BIBentrySTDinterwordspacing

\bibitem{polese22understanding}
\BIBentryALTinterwordspacing
M.~Polese, L.~Bonati, S.~D'Oro, S.~Basagni, and T.~Melodia, ``Understanding
  {O-RAN:} architecture, interfaces, algorithms, security, and research
  challenges,'' \emph{CoRR}, vol. abs/2202.01032, Oct. 2022. [Online].
  Available: \url{https://arxiv.org/abs/2202.01032}
\BIBentrySTDinterwordspacing

\bibitem{zhang18loadbalancing}
H.~{Zhang}, L.~{Song}, and Y.~J. {Zhang}, ``Load balancing for 5{G} ultra-dense
  networks using device-to-device communications,'' \emph{{IEEE} Trans.
  Wireless Commun.}, vol.~17, no.~6, pp. 4039--4050, April 2018.

\bibitem{Handover_Wang18}
Z.~{Wang}, L.~{Li}, Y.~{Xu}, H.~{Tian}, and S.~{Cui}, ``Handover control in
  wireless systems via asynchronous multiuser deep reinforcement learning,''
  \emph{{IEEE} Internet Things J.}, vol.~5, no.~6, pp. 4296--4307, Dec. 2018.

\bibitem{chinchali18cellular}
S.~Chinchali \emph{et~al.}, ``Cellular network traffic scheduling with deep
  reinforcement learning,'' in \emph{Proc. AAAI}, Apr. 2018, pp. 1--6.

\bibitem{zhao19deep}
N.~Zhao, Y.-C. Liang, D.~Niyato, Y.~Pei, M.~Wu, and Y.~Jiang, ``Deep
  reinforcement learning for user association and resource allocation in
  heterogeneous cellular networks,'' \emph{{IEEE} Trans. Wireless Commun.},
  vol.~18, no.~11, pp. 5141--5152, Nov. 2019.

\bibitem{xu19load}
Y.~Xu, W.~Xu, Z.~Wang, J.~Lin, and S.~Cui, ``Load balancing for ultradense
  networks: A deep reinforcement learning-based approach,'' \emph{{IEEE}
  Internet Things J.}, vol.~6, no.~6, pp. 9399--9412, 2019.

\bibitem{gupta21load}
M.~Gupta \emph{et~al.}, ``Load balancing and handover optimization in
  multi-band networks using deep reinforcement learning,'' in \emph{Proc. IEEE
  GLOBECOM}, Dec. 2021, pp. 1--6.

\bibitem{wu21load}
D.~Wu \emph{et~al.}, ``Load balancing for communication networks via
  data-efficient deep reinforcement learning,'' in \emph{Proc. IEEE GLOBECOM},
  Dec. 2021, pp. 01--07.

\bibitem{lacava22programmable}
\BIBentryALTinterwordspacing
A.~Lacava \emph{et~al.}, ``Programmable and customized intelligence for traffic
  steering in {5G} networks using open {RAN} architectures,'' \emph{CoRR}, vol.
  abs/2209.14171, Oct. 2022. [Online]. Available:
  \url{https://arxiv.org/abs/2209.14171}
\BIBentrySTDinterwordspacing

\bibitem{mai2022sample}
V.~Mai, K.~Mani, and L.~Paull, ``Sample efficient deep reinforcement learning
  via uncertainty estimation,'' in \emph{Proc. ICLR}, Apr. 2022, pp. 1--6.

\bibitem{alcaraz22model}
J.~J. Alcaraz, F.~Losilla, A.~Zanella, and M.~Zorzi, ``Model-based
  reinforcement learning with kernels for resource allocation in {RAN}
  slices,'' \emph{{IEEE} Trans. Wireless Commun.}, Sept. 2022, (early access).

\bibitem{agarwal2022taskdriven}
S.~Agarwal and S.~P. Chinchali, ``Task-driven data augmentation for
  vision-based robotic control,'' in \emph{Proc. CoRL}, Dec. 2022.

\bibitem{vishwanath02opportunistic}
P.~Viswanath, D.~Tse, and R.~Laroia, ``Opportunistic beamforming using dumb
  antennas,'' \emph{{IEEE} Trans. Inf. Theory}, vol.~48, no.~6, pp. 1277--1294,
  June 2002.

\bibitem{yigal07fairness}
Y.~Bejerano, S.-J. Han, and L.~Li, ``Fairness and load balancing in wireless
  {LANs} using association control,'' \emph{IEEE/ACM Trans. on Netw.}, vol.~15,
  no.~3, pp. 560--573, June 2007.

\bibitem{gupta22system}
M.~Gupta, I.~P. Roberts, and J.~G. Andrews, ``System-level analysis of
  full-duplex self-backhauled millimeter wave networks,'' \emph{{IEEE} Trans.
  Wireless Commun.}, Sept. 2022, (early access).

\bibitem{rasek20joint}
M.~E. Rasekh, D.~Guo, and U.~Madhow, ``Joint routing and resource allocation
  for millimeter wave picocellular backhaul,'' \emph{{IEEE} Trans. Wireless
  Commun.}, vol.~19, no.~2, pp. 783--794, Feb. 2020.

\bibitem{srikant2014comnets}
R.~Srikant and L.~Ying, \emph{Communication Networks: An Optimization, Control
  and Stochastic Networks Perspective}.\hskip 1em plus 0.5em minus 0.4em\relax
  New York, NY, USA: Cambridge University Press, 2014.

\bibitem{mpc_james}
J.~B. Rawlings, D.~Q. Mayne, and M.~M. Diehl, \emph{Model Predictive Control:
  Theory, Computation, and Design}.\hskip 1em plus 0.5em minus 0.4em\relax
  Santa Barbara, CA, USA: Nob Hill Publishing, 2017.

\bibitem{liang03predictive}
B.~Liang and Z.~Haas, ``Predictive distance-based mobility management for
  multidimensional {PCS} networks,'' \emph{IEEE/ACM Trans. on Netw.}, vol.~11,
  no.~5, pp. 718--732, Oct. 2003.

\bibitem{tayyab19survey}
M.~Tayyab, X.~Gelabert, and R.~Jäntti, ``A survey on handover management: From
  {LTE} to {NR},'' \emph{IEEE Access}, vol.~7, pp. 118\,907--118\,930, Aug.
  2019.

\bibitem{cvxpy1}
S.~Diamond and S.~Boyd, ``{CVXPY}: {A} {P}ython-embedded modeling language for
  convex optimization,'' \emph{J. Mach. Learn. Res.}, vol.~17, no.~83, pp.
  1--5, Jan. 2016.

\bibitem{wang13theoretical}
Y.~Wang, L.~Wang, Y.~Li, D.~He, and T.-Y. Liu, ``A theoretical analysis of
  {NDCG} type ranking measures,'' in \emph{Proc. Mach. Learn. Res.},
  vol.~30.\hskip 1em plus 0.5em minus 0.4em\relax PMLR, June 2013, pp. 25--54.

\bibitem{pobrotyn2021neuralndcg}
P.~Pobrotyn and R.~Bialobrzeski, ``{NeuralNDCG}: Direct optimisation of a
  ranking metric via differentiable relaxation of sorting,'' \emph{ArXiv}, vol.
  abs/2102.07831, 2021.

\bibitem{jaggi13revisiting}
M.~Jaggi, ``Revisiting {Frank-Wolfe}: Projection-free sparse convex
  optimization,'' in \emph{Proc. ICML}, vol.~28, June 2013, pp. 427--435.

\bibitem{pedregosa2020linearly}
F.~Pedregosa, G.~Negiar, A.~Askari, and M.~Jaggi, ``Linearly convergent
  {Frank-Wolfe} with backtracking line-search,'' in \emph{Proc. AISTATS}, Mar.
  2020, pp. 1--6.

\bibitem{3GPP2017}
\emph{{Study on channel model for frequencies from 0.5 to 100 GHz ({R}elease
  14)}}, document 3GPP TR 38.901, Sep. 2017.

\bibitem{3GPP_RRC_2020}
\emph{{Evolved Universal Terrestrial Radio Access (E-UTRA); Radio Resource
  Control (RRC); Protocol specification}}, document 3GPP TS 36.331, Sept. 2020.

\bibitem{schulman17proximal}
\BIBentryALTinterwordspacing
J.~Schulman, F.~Wolski, P.~Dhariwal, A.~Radford, and O.~Klimov, ``Proximal
  policy optimization algorithms,'' \emph{CoRR}, vol. abs/1707.06347, June
  2017. [Online]. Available: \url{https://arxiv.org/abs/1707.06347}
\BIBentrySTDinterwordspacing

\end{thebibliography}
\end{document}